# Emergent magnetism with continuous control in the ultrahigh conductivity layered oxide PdCoO$_2$

Matthew Brahlek[1*], Alessandro R. Mazza[1,2], Abdulgani Annaberdiyev[3], Michael Chilcote[1], Gaurab Rimal[4], Gábor B. Halász[1], Anh Pham[3], Yun-Yi Pai[1], Jaron T. Krogel[1], Jason Lapano[1], Benjamin J. Lawrie[1], Gyula Eres[1], Jessica McChesney[5], Thomas Prokscha[6], Andreas Suter[6], Seongshik Oh[4], John W. Freeland[5], Yue Cao[7], Jason S. Gardner[1], Zaher Salman[6], Robert G. Moore[1], Panchapakesan Ganesh[3#], T. Zac Ward[1]

[1]Materials Science and Technology Division, Oak Ridge National Laboratory, Oak Ridge, TN, 37831, USA
[2]Center for Integrated Nanotechnologies, Los Alamos National Laboratory, Los Alamos, NM
[3]Center for Nanophase Materials Sciences, Oak Ridge National Laboratory, Oak Ridge, TN, 37831, USA
[4]Department of Physics and Astronomy, Rutgers, The State University of New Jersey, Piscataway, NJ, 08854, USA
[5]Advanced Photon Source, Argonne National Laboratory, Lemont, IL, 60439, USA
[6]Laboratory for Muon Spin Spectroscopy, Paul Scherrer Institute, CH-5232 Villigen PSI, Switzerland
[7]Materials Science Division, Argonne National Laboratory, Lemont, IL, 60439, USA
Correspondence should be addressed to *brahlekm@ornl.gov, #ganeshp@ornl.gov

**Abstract:** The current challenge to realizing continuously tunable magnetism lies in our inability to systematically change properties such as valence, spin, and orbital degrees of freedom as well as crystallographic geometry. Here, we demonstrate that ferromagnetism can be externally turned on with the application of low-energy helium implantation and subsequently erased and returned to the pristine state via annealing. This high level of continuous control is made possible by targeting magnetic metastability in the ultra-high conductivity, non-magnetic layered oxide PdCoO$_2$ where local lattice distortions generated by helium implantation induce emergence of a net moment on the surrounding transition metal octahedral sites. These highly-localized moments communicate through the itinerant metal states which triggers the onset of percolated long-range ferromagnetism. The ability to continuously tune competing interactions enables tailoring precise magnetic and magnetotransport responses in an ultra-high conductivity film and will be critical to applications across spintronics.

Designing novel materials systems with flexible control over magnetic phase and properties is central to enabling new devices with superior energy efficiency and increased computational power(*1*, *2*). Specifically, magnetic metals are critical to the development of next-generation spintronic devices based on, for example, spin-transfer torques. Yet, tailoring magnetic properties in traditional systems has been limited to controlling interfaces or alloying(*3*). Tuning magnetism in a continuous way is challenging, since the electronic and magnetic character are linked through the exchange interaction to the specific intrinsic properties within a material, including their charge states, net moments, and bonding environments. Targeted control of these interactions has been achieved via external voltage control(*4*), thickness-scaling(*5–7*) and strain in freestanding films(*8*) in systems that are already magnetic, yet new routes are necessary for future applications where magnetism can be induced and controlled in a continuous and metastable way. Towards this, transition metal oxides possess strongly correlated spin, orbital, and charge degrees of freedom all of which are necessary for materials tunablility(*9*), but, in practice, magnetic metal oxides are rare and usually exhibit relatively high resistivities(*10*, *11*). Notable exceptions are the 2-dimensional (2D) *AB*O$_2$ metallic delafossites, shown in Fig.1a, which represent a unique material system that has garnered significant recent attention(*12*, *13*) due to a myriad of unusual fundamental properties(*14–17*) as well as novel functional responses(*18–21*). For PdCoO$_2$, PtCoO$_2$ and PdCrO$_2$, this class of oxides provides a highly anisotropic layered crystal structure in which the *A*-site and *B*-site sublattice layers have extraordinarily different properties(*12*). Specifically, the Pd/Pt-layers give rise to extremely high conductivity(*22*), whereas the states in the Co/Cr layers are highly localized and insulating(*23*, *24*). The sole magnetic delafossite, Cr-based PdCrO$_2$ exhibits a stable and non-tunable antiferromagnetic state(*25*). On the other hand, Co-based PdCoO$_2$ is a non-magnetic, Pauli paramagnet but likely resides near a magnetic metastability. Here, the octahedrally coordinated Co$^{3+}$, shown in Fig.1b, is in a low-spin state that is known to be in proximity to the intermediate (*M*=2$\mu_B$/Co) and high (*M*=4$\mu_B$/Co) spin states accessible through strain and charge doping(*26*). Similarly, bulk Pd is also well-known to be close to a Stoner instability due to the high density of states at the Fermi level ($E_F$)(*27*), which can become magnetic when made atomically thin(*28*) or at an interface with a magnet(*29*), as shown in Fig.1c. Moreover, the strong spin-orbit coupling associated with metal layers of Pd can give rise to important functional properties, such as magnetic anisotropy, spin torque, novel domain wall structures, and topological spin textures(*1*). While attempts at stabilizing ferromagnetism using traditional magnetic doping on the Co-site have not been successful(*30*), hints of surface magnetism have been observed in cleaved bulk crystals(*31, 32*) and in thin films(*33*). A systematic uniform modification of the bulk-like magnetic responses in PdCoO$_2$ have not been predicted nor demonstrated experimentally, which highlights that new routes are needed to induce and control this fundamental property. These aspects together make PdCoO$_2$ an extremely promising candidate for tailoring magnetism in a high-conductivity, low-dimensional/nanostructured platform.

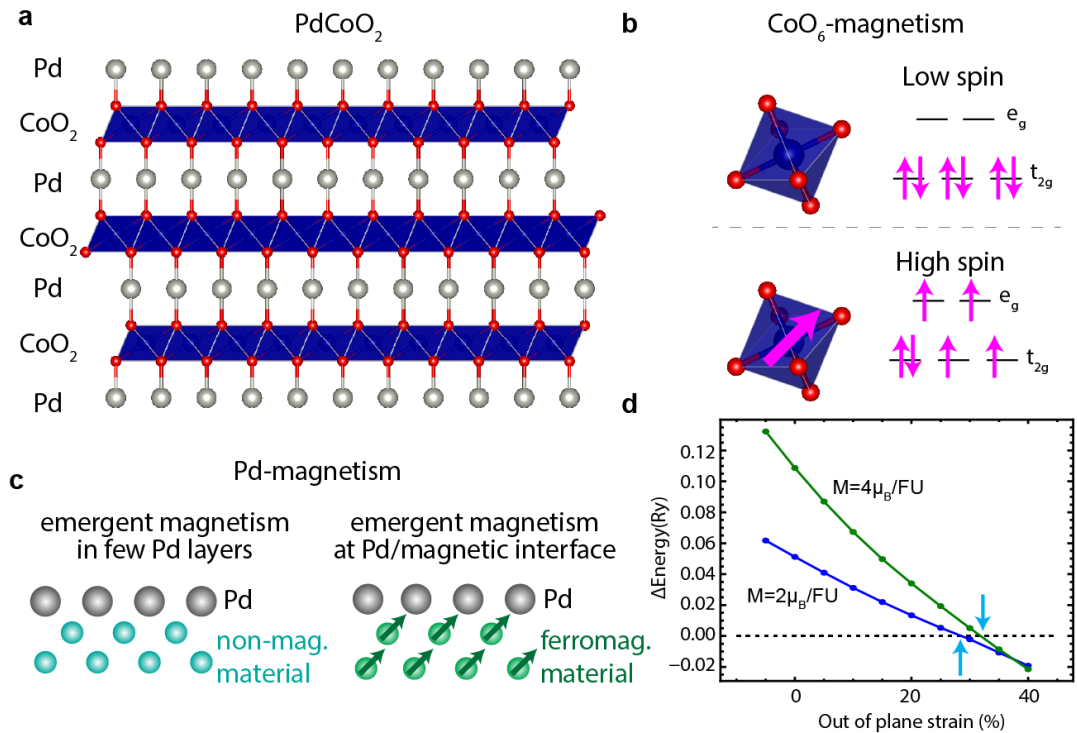

**Fig. 1. Illustration of magnetic metastability of PdCoO$_2$.** (**a**) Schematic of the delafossite structure composed of alternating Pd and CoO$_2$ layers. (**b-c**) Magnetic metastability in PdCoO$_2$ due to the close proximity of the non-magnetic ($M=0\mu_B$/Co low-spin) and a magnetic state ($M=4\mu_B$/Co high-spin, shown) known in octahedral coordinated Co$^{3+}$, and c the large density of states of bulk Pd, which is known to become magnetic in the monolayer limit (left), and when interfaced with a ferromagnetic material (right), represented by green arrows. (**d**) Calculated energy difference among the strained magnetic cells with intermediate ($M=2\mu_B$/formula unit (FU), blue) and high-spin ($M=4\mu_B$/FU, green) relative to the unstrained non-magnetic states, which shows a crossover with a large out-of-plane strain (indicated by the blue arrows).

Here, we reveal a route to induce magnetism with a high level of continuous control in highly conducting PdCoO$_2$ thin films. Experimentally it is shown that the helium implantation in PdCoO$_2$ can systematically cause strain along the out-of-plane direction, while fully retaining the structural and chemical character. This is effectively found to tune from the non-magnetic ground state of the pristine structure to a paramagnetic state with net moments and finally to a ferromagnetic ground state that exhibits long range order and clear hysteresis with onset near 150 K. These results closely match models in which uniaxial out-of-plane strain pushes the Co sites into a spin-polarized state near local lattice distortions; the induced moments on the Co site then magnetize the itinerant Pd states to stabilize long-range ferromagnetic order. Magnetotransport is used to map the evolution of the anomalous Hall effect (AHE), which clearly demonstrates a strain-induced crossover in scattering mechanisms. Finally, the magnetic state is found to be fully reversible via high temperature annealing, which resets the system to the pristine state. These observations give insight into how magnetism emerges in a low-dimensional system that combines both itinerant and localized physics, as well as being a highly tunable platform for creating nanoscale magnetic devices with previously inaccessibly high electrical conductivity.

To understand how strain can be used to induce magnetism in $PdCoO_2$, first-principles calculations were performed as a function of distortion to the unit cell (Supporting Information Section I). As previous reports have shown, helium implantation can be used to generate an out-of-plane expansion of the lattice while the in-plane lattice remains heteroepitaxially locked and unchanged (Refs.(*34–37*)), so calculations were performed by applying an out-of-plane expansion to the unit cell. Since $PdCoO_2$ shows itinerate and localized physics, density functional theory + U calculations(*38*) informed by quantum Monte Carlo (QMC)(*39, 40*) were performed to minimize ambiguities in the choice of the *U* parameter(*41*). Figure 1d shows the energy difference for the distorted unit cell with $M=2\mu_B$ per formula unit (FU, the spin is not confined to the Co, see Supporting Information; intermediate spin) and $M=4\mu_B$/FU (high spin) relative to the undistorted, non-magnetic unit cell versus out-of-plane strain. With increasing strain, there is a crossover into a magnetic phase near 30%, as marked by the blue arrows. Moreover, there is a large mixing of the magnetic Co into the density of states at the Fermi level (see Fig.S5). This mixing of magnetic Co orbital character into the itinerant states generates a large hybridization between the Pd and Co electrons and is central to $PdCoO_2$ becoming magnetic. As such, novel approaches that are capable of creating and inducing magnetism in the delafossites are needed since progress towards percent-level epitaxial strain(*13, 42–44*) and magnetic-doping-control in the delafossites(*30*) have been challenging. Metastable helium implantation is ideal to achieve very large distortions(*34–36*) that can locally reach the large, 30% strain that is necessary to drive $PdCoO_2$ into a magnetic state.

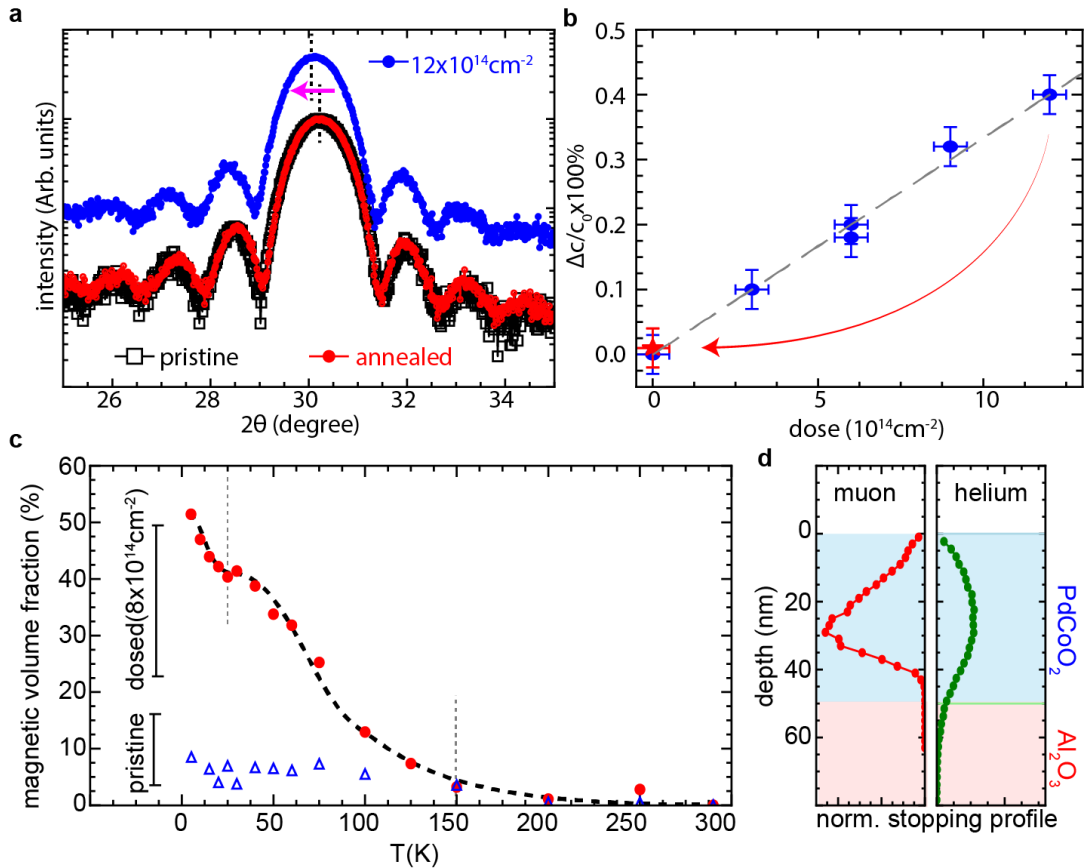

**Fig. 2. Implantation induced strain and emergent ferromagnetism.** (**a**) $2\theta$-$\theta$ scans about the 006 $PdCoO_2$ reflection for a pristine sample, an implanted sample, and a sample reset with a post implantation anneal. (**b**) Systematic expansion of the out-of-plane lattice parameter, $\Delta c/c_0 \times 100\%$, for an implanted series (blue circles) and a sample what was reset with a post implantation anneal (red star). (**c**) Temperature dependence of the magnetic volume fraction for an implanted sample and pristine sample measured with low-energy muon spin rotation. (**d**) Simulated normalized muon implantation profile, (left panel) in comparison to the helium implantation profile (right panel).

To test if distortions of the unit cell produce magnetism, detailed structural and electronic characterization were performed as a function of helium dose on a series of molecular beam epitaxy grown thin films (thickness ≈10 nm, see Ref.(*44*) and Supporting Information Section IIA for growth and implantation details). Figure 2a shows X-ray diffraction $2\theta$-$\theta$ scans for the 006 reflection for pristine and highly dosed $PdCoO_2$ samples. Here, implantation rigidly shifts the peak to lower $2\theta$, which corresponds to a coherent out-of-plane expansion. The change in the out-of-plane lattice parameter $c$, shown by $\Delta c/c \times 100\%$ in Fig.2b, is linear with dose and gives a maximum average strain of ~0.5% at a dose of $12\times10^{14}$ cm$^{-2}$. Although the net or average strain is much less than that predicted by theory (~30%), there are likely regions of large strain surrounding an implanted helium atom, which we show later produces local moments. Moreover, the overall peak shape is unchanged, indicating that the dosing is homogeneous and structural integrity is maintained. In a wide range $2\theta$-$\theta$ scan (Fig.S8), no additional peaks arise, which indicates that there is no damage to the film; for example, no Pd, Co, or PdCo peaks are observed(*19*, *20*). Secondly, the small Laue oscillations about the main peak indicate that the films are nearly atomically flat, which, importantly, is not changed during implantation. This indicates that the surfaces are not damaged, nor is the top layer sputtered away, thereby thinning the sample. This result is corroborated by the x-ray reflectivity measurements shown in Fig.S9 and x-ray photoemission spectroscopy shown in Fig.S10.

We highlight that the implantation process can be fully reset via annealing: Fig.2a shows a $2\theta$-$\theta$ scan for a $PdCoO_2$ film that was implanted then annealed at 750 °C in air (red symbols), which is overlaid on the pristine data. The peak position shows that the c-axis lattice parameter exactly returns to the pristine state (red stars). Additionally, the peak character and Laue oscillations demonstrate that there is no change in the crystalline thickness or lattice quality that would occur if surface sputtering or bulk amorphousizing were present. Throughout these experiments, we saw no sign of damage or sample degradation across multiple implant/anneal steps. In contrast to previous studies(*45–47*), this confirms that all effects induced by helium implantation are both stable to thermal processing as well as fully reversible.

To show that changes to lattice volume induces magnetism, low-energy muon spin rotation (μSR) measurements were performed, which is an extremely sensitive and direct probe of the depth-resolved magnetic profile in thin films (see Ref.(*48*, *49*) and Supporting Information Section-IID and IIIB). The magnetic volume fraction (MVF), extracted from fitting the muon decay profile, is shown for muon implantation energy of 6 keV for the dosed film as well as a pristine film. Here, a non-zero MVF indicates that magnetism can be created and stabilized using implantation, as shown in Fig.2c for a film dosed at $8\times10^{14}$ cm$^{-2}$ (circles). For this incident energy the muons probe deep into the bulk of the film, as shown in Fig.2d for the implantation distribution calculated using TRIM.SP for muons(*50*, *51*) in comparison to the helium dosed profile calculated with SRIM(*52*). The MVF is found to turn on below 150 K, as marked by the dashed gray line, and steadily grows as the temperature is reduced. In addition to the high-temperature turn-on, there is a transition at 25 K with an upturn in the MVF also marked by a dashed gray line. The MVF peaks at around 55%, which is expected for the inhomogeneous profiles of both the helium and the muons, which probe simultaneously optimally and underdosed regions of the film, as discussed later for the anomalous Hall effect. Moreover, in pristine films there is a weak magnetic signature, which likely originates from native oxygen vacancies on the surface(*33*), as discussed below. The key here is that there is a strong magnetic signature in the bulk of the film. This shows that the distortions produced by the helium causes bulk magnetism. This effect cannot be associated with native oxygen vacancies since it is significantly higher in MVF than the pristine films. Altogether, muon spectroscopy conclusively demonstrates that the helium implantation process can be used to induce stable bulk magnetism in otherwise non-magnetic $PdCoO_2$.

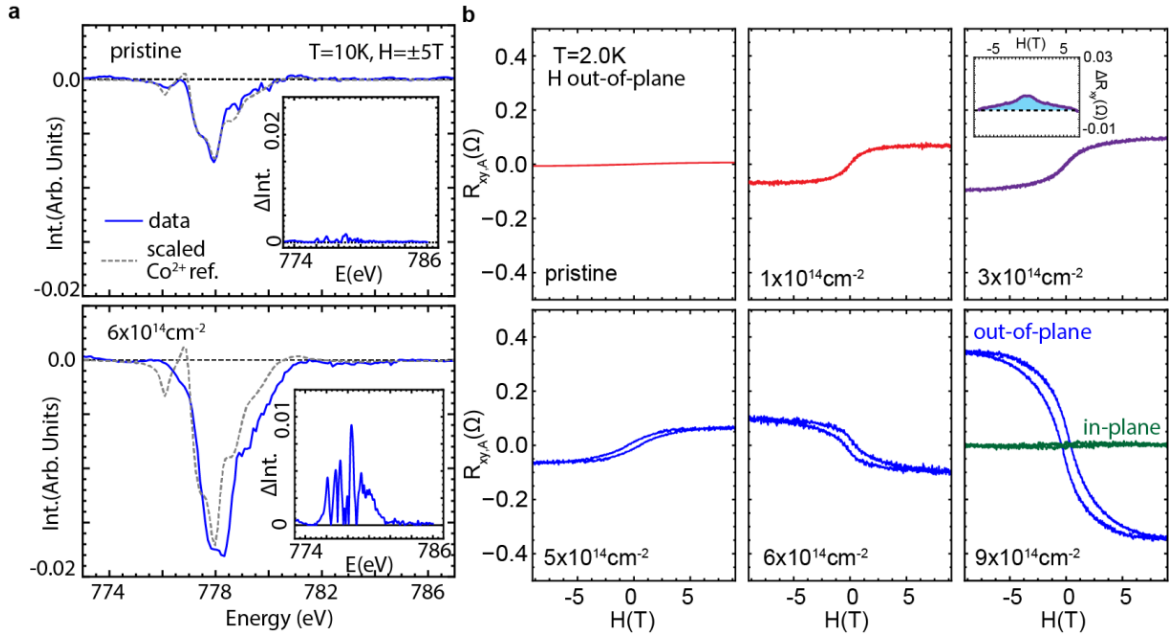

**Fig. 3. Continuously controlled magnetism and anomalous Hall effect.** (**a**) XMCD spectra for a pristine (top) and implanted (bottom) sample performed in $H=5$ T and $T=10$ K. Insets, difference of measured spectra relative to a $Co^{2+}$ reference spectrum. $Co^{2+}$ data was taken from Ref. (*53*). (**b**) Anomalous Hall resistance versus magnetic field with increasing dose, as labeled, with the temperature fixed at 2 K and the magnetic field orientated out-of-plane for all panels, and also in-plane for the highest dose ($9\times10^{14}$ cm$^{-2}$, right). Inset shows the difference between the anomalous Hall resistance taken on the up-sweep and down-sweep, $\Delta R_{xy,A}$.

This emergent magnetism is further interrogated using x-ray magnetic circular dichroism (XMCD) to observe element-specific charge and spin state evolutions with ion implantation. Figure 3a shows the XMCD signal for a pristine sample (top) and a sample dosed into the ferromagnetic regime ($6\times10^{14}$ cm$^{-2}$, bottom) versus photon energy as solid blue curves, which were performed at the Co $L_3$ edge in a magnetic field of $\pm5$T and a temperature of 10 K. First, the pristine sample shows a clear non-zero XMCD signal, indicating the presence of moments on the Co site. There is an excellent agreement between the reference spectrum for $Co^{2+}$ (shown in gray; from Ref.(53)) and the pristine sample (shown in blue), which indicates percent-level oxygen vacancies. This can be seen more clearly in the inset where the reference spectrum was subtracted from the pristine $PdCoO_2$ spectrum. This is consistent with oxygen vacancies being the origin for the non-zero MVF shown for the pristine film in Fig.2c. In contrast, the implanted sample shows a larger (~2-3×) XMCD signal with a significantly different line shape, which can be seen by comparison with the same $Co^{2+}$ reference in the main panel and inset. The change in line shape indicates a clear deviation from $Co^{2+}$, which is consistent with a modification of the $Co^{3+}$ spin-state(54). This shows that the magnetic state induced with implantation is different than the magnetic character of oxygen vacancies related to magnetic $Co^{2+}$, which is confirmed next by contrasting the dramatic difference in Hall response for pristine vs $6\times10^{14}$ cm$^{-2}$. Based on the percent-level of magnetic Co via XMCD, the length scale of the magnetic interaction must be on the order of nanometers near the paramagnetic-ferromagnetic crossover. This is consistent with the magnetic interaction being long-range in nature and mediated by the itinerate electrons in the Pd layer via, for example, the RKKY mechanism(55), which are dominated by the itinerate Pd states.

The AHE is an extremely sensitive indicator of a material's underlying magnetic and electronic properties, as it stems from both the internal magnetization and the properties of the conduction electrons (i.e., band structure and scattering processes; see (56) and Supporting Information Section-III). In Fig.3b, we show the anomalous Hall resistance, $R_{xy,A}$, versus magnetic field at a temperature of 2 K for an implantation series with increasing dose, as labeled. For the pristine sample, $R_{xy,A}$ is nominally flat with magnetic field, which indicates that there is very little magnetic response. This is consistent with non-magnetic $PdCoO_2$, with the small background likely due to the dilute Co moments seen in XMCD (seen also in Ref.(30)). With a small dosage, $1\times10^{14}$ cm$^{-2}$–$2\times10^{14}$ cm$^{-2}$, $R_{xy,A}$ becomes dramatically non-zero, indicating the emergence of magnetism. The character of the AHE is significantly different for $1\times10^{14}$ cm$^{-2}$ compared to the pristine sample, demonstrating a minor role of native oxygen vacancies. Here, $R_{xy,A}$ shows an *S*-shape with $R_{xy,A}(H=-9T)<0$, and $R_{xy,A}(H=9T)>0$. For the low dosage, no hysteresis is observed down to 40 mK (see Fig.S11), which shows that net moments have been created yet do not order, i.e., the material consists of a dynamic, disordered spin system. For $3\times10^{14}$ cm$^{-2}$, the shape of $R_{xy,A}$ vs. *H* is slightly flatter, and there is discernible hysteresis at zero magnetic field (inset Fig.3b), which vanishes at a Curie temperature of $T_C \approx 5\pm 2$ K. The hysteresis indicates ferromagnetic ordering, consistent with muon spectroscopy results shown in Fig.2c. At $5\times 10^{14}$ cm$^{-2}$ the shape of $R_{xy,A}$ flattens and the zero field hysteresis becomes more prominent. For dosages of $6 \times 10^{14}$ cm$^{-2}$–$9\times 10^{14}$ cm$^{-2}$, $R_{xy,A}(9T)$ surprisingly shows a complete reversal of sign and grows in magnitude with hysteresis closing at a maximal $T_C$ of around $25\pm 2$ K. Moreover, the saturation field is systematically pushed to $>9$ T. Finally, for $9\times 10^{14}$ cm$^{-2}$, this measurement was extended to in-plane fields in which case it shows a nearly flat AHE. The pronounced difference between in-plane and out-of-plane fields indicates that the emergent ferromagnetic state likely exhibits strong perpendicular magnetic anisotropy with the easy axis pointed out-of-plane. Since the film is thin (10 nm), shape anisotropy should favor an in-plane easy axis, which is quite surprising and is of considerable technological relevance(*1*), for example, for spin logic and memory technologies based on spin-transfer torques(*57*).

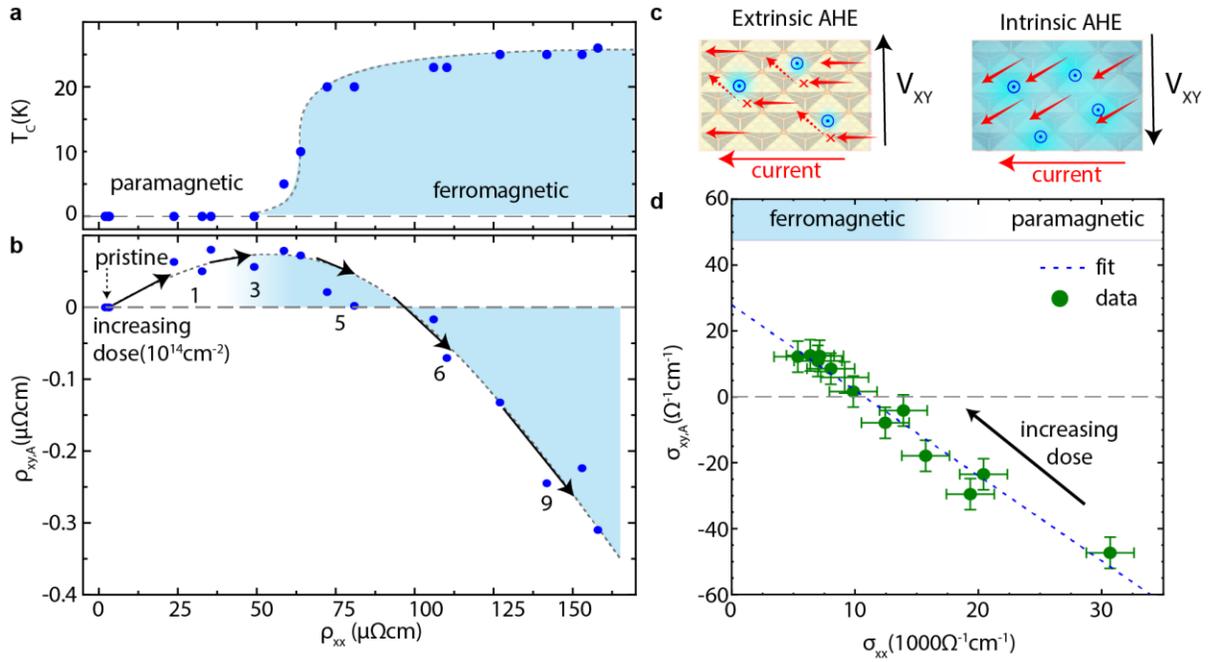

**Fig. 4. Tailored evolution of competing anomalous Hall effect mechanisms.** (**a**) Ferromagnetic transition temperature versus the increase in the in-plane resistance. $T_C$ was obtained from the data in Fig. 3 at which the hysteresis vanishes. (**b**) Anomalous Hall resistance taken at $H=9T$ plotted versus in-plane resistance where select dosages are indicated in the panel. (**c**) Schematic of the possible mechanisms that contribute to the anomalous Hall resistance when a current is applied towards the left in a magnetic field oriented out of the page. (left) Extrinsic scattering process off magnetic sites (blue dotted-circles) in a paramagnet (net yellow background) causes deflection orthogonal to the direction of the applied current (dashed arrows). (right) Intrinsic mechanism results in a non-zero velocity of carriers orthogonal to the direction of the applied current in a ferromagnet (net blue background). Though the sign may vary, both processes cause an accumulation of charge on one side (top or bottom) of the material, resulting in a non-zero Hall voltage ($V_{xy}$). (**d**) Anomalous Hall data in b replotted in terms of conductivities. This follows a clear linear trend, and the corresponding fit is indicated as the dashed blue curve.

The complete data set characterizing the magnetic evolution with helium-implantation-induced strain is shown in Fig.4, where $T_C$ (extracted from the onset of hysteresis) and the anomalous Hall resistivity, $\rho_{xy,A}(9T)=R_{xy,A}(9T)\times thickness$ are plotted versus resistivity, $\rho_{xx}=R_{xx}(9T)\times thickness$, at 2 K (it is noted $R_{xx}$ has small field dependence). From this we see that dosed $PdCoO_2$ maintains an extremely low resistivity, especially when compared to typical oxides and magnetic oxides(*5, 6, 11*) (see Fig.S13 for a plot of $\rho_{xx}$ versus dose). Furthermore, the plot of $T_C$ shows the critical dosage where ferromagnetism onsets with $T_C$ quickly increasing from 0 to ~25 K and then saturating. Ferromagnetic ordering exactly coincides with the lower temperature transition in the MVF shown in Fig.2c. In Fig.4b, $\rho_{xy,A}(9T)$ is stacked vertically below $T_C$ to highlight the evolution of the AHE as magnetic moments arise and then subsequently form long-range order. Here, in the paramagnetic state, $\rho_{xy,A}(9T)$ increases roughly linearly with increasing resistivity. As ferromagnetism is established (>$3\times10^{14}$ cm$^{-2}$), $\rho_{xy,A}(9T)$ flattens with increasing $\rho_{xx}$, then $\rho_{xy,A}(9T)$ starts to decrease slightly faster than linearly. The coincidence of the flattening of $\rho_{xy,A}$ with the onset of ferromagnetism and the subsequent change in sign of $\rho_{xy,A}$ indicates that there are two clear competing mechanisms of the AHE, which can be finely controlled with helium implantation.

AHE arises from several mechanisms defined by specific resistivity regimes(*56*). For the lowest resistivity regime ($\rho_{xx}$<1μΩ·cm), the AHE is dominated by extrinsic mechanisms, for example, skew-scattering, Fig.4c left panel. For higher resistivities (1μΩ·cm<$\rho_{xx}$<100μΩ·cm), intrinsic effects dominate the AHE. As schematically shown in Fig.4c right panel, the Karplus-Luttinger effect occurs when carriers obtain a component of the Fermi velocity orthogonal to the current, which is linked to the band structure in the context of the Berry phase. A comparison with theory can be made by plotting the anomalous Hall conductivity $\sigma_{xy,A}$ versus the longitudinal conductivity $\sigma_{xx}$. Here, theory predicts $\sigma_{xy,A}\propto\sigma_{xx}^n$, with $n$=1 in the extrinsic regime and $n$=0 in the intrinsic regime(*56*). The plot of $\sigma_{xy,A}$ versus $\sigma_{xx}$ is shown in Fig.4d. The data shows a very clear linear scaling ($n$=1). If the anomalous contribution to the anomalous Hall conductivity are additive, then $\sigma_{xy,A}=\sigma_{xx,I}+\sigma_{xx,E}$. Here, $\sigma_{xx,I}$ is the intrinsic contribution, independent of $\sigma_{xx}$, and $\sigma_{xx,E}$ is the extrinsic contribution, linearly dependent on $\sigma_{xx}$ (i.e. $n$=1). This yields $\sigma_{xy,A}=\sigma_{xx,I}+\alpha\times\sigma_{xx}$ which was used to fit the data in Fig.4(d) and shows excellent agreement. The fit crosses $\sigma_{xy,A}=0$ around $\sigma_{xx}=10.7\times10^4$ $\Omega^{-1}$cm$^{-1}$ and extrapolating to $\sigma_{xx}\to 0$ gives $\sigma_{xy,I}=28$ $\Omega^{-1}$cm$^{-1}$. In contrast to expectations(*56*), we find no well-defined crossover regime, but rather that intrinsic and extrinsic effects are additive. It is quite interesting that skew scattering continues to be active deep into the low conductivity regime. Specifically, the intrinsic AHE comes from anomalies in the band structure. Low-energy scattering processes (extrinsic effects such as skew scattering) are highly dependent on the electronic properties near the Fermi level which may significantly change across this regime. The fact these contributions continue to be additive may point out that extrinsic effects are not disrupted by changes to the electronic structure even deep into the intrinsic regime. This is likely related to the dilute density of magnetic sites where magnetism is mediated across long length scales due to the unique, highly itinerate Pd states. Furthermore, the extrapolation to $\sigma_{xx}\to 0$ enables direct identification of the strength of both the intrinsic and extrinsic contributions. The anomalous Hall conductivity in implanted PdCoO$_2$ is lower than typical magnetic metals, Fe, Co, Ni, Gd, Fe-Co/Si alloys, Mn$_3$Sn, Mn$_3$Ge, MnGe, Co$_3$Sn$_2$S$_2$, KV$_3$Sb$_5$ and Fe$_3$Sn$_2$, and oxides, SrRuO$_3$ and (La,Sr)CoO$_3$, where $\sigma_{xy,A}\approx 100$-$1000$ $\Omega^{-1}$cm$^{-1}$ (see Ref.(*56*, *58*, *59*) and references therein). Given the dilute density of magnetic sites in implanted PdCoO$_2$ based on XMCD, $\sigma_{xy,I}\approx 28$ $\Omega^{-1}$cm$^{-1}$ is extremely large per magnetic site, which may result from the 2D nature and the strong spin-orbit coupling in the Pd layer. The ability to use helium to continuously tune magnetism across these regimes provides new insight into the fundamental mechanisms and gives a route to control and design a material with a desired response.

The strong hybridization among magnetic Co sites and subsequent polarization of the strong spin-orbit coupled Pd layer is central to the large, tailorable magnetic response observed. Yet, the simple distortion-induced picture regarding the CoO layer and the interplay with the itinerate Pd states that formed our hypothesis is likely an oversimplification and many theoretical and experimental questions arise regarding the origins of the emergent magnetism. This diversity and richness of the implantation process offers many routes to design magnetism for specific applications. Gaining this control requires a deeper understanding of how the implantation process can be used to target specific defect types. Given that the magnetic ordering onset is observed at 150 K, exploring the detailed magnetic properties of each type of defect and how to target them is an important critical future question to boost $T_C$ towards room temperature.

Moreover, the use of helium to dynamically induce magnetism can be applicable beyond just PdCoO$_2$ and should be generally usable to tailor magnetism in the other Pd/Pt-based delafossites(*12*), at magnetic oxide interfaces with Pd/Pt, as well as 2D layered magnetic materials(*60*).

In contrast to adjusting preexisting magnetic properties of metals with nanometer scale resolution using ion technologies(*46*, *47*, *61*) or micrometer scale resolution via thermal induced optical pulses(*45*), the key novelty is that it is possible to induce emergent and stable ferromagnetism in the otherwise non-magnetic oxide material using helium implantation. This generally applicable approach solves a critical challenge for the design of magnetic responses in materials with strong spin-orbit coupling and high conductivities. Here, the magnetism is quite stable, which enables devices to be written and then erased with a high-temperature anneal. Therefore, there are many possible routes for the creation of magnetic devices by precisely controlling the implantation profile across the depth but also with nm-scale lateral resolution(*47*). Specifically, varying the implantation dose across the thickness can enable the creation of vertically stacked multilayer magnetic device where each layer has a specific magnetic property, laterally structured magnonic crystals, or arrays of technologies where high-conductivity, low-dissipation interconnects can be based on pristine PdCoO$_2$. As this is a fully tunable magnetic system, helium-implanted PdCoO$_2$ is a new platform to explore the interplay among local/itinerate and extrinsic/intrinsic effects that will have a profound impact on the fundamental understanding of magnetic responses and open many routes for technological adaptation.

**Supporting Information**
Supporting Information: In Section I, Ab-initio DFT and QMC Supporting Data, Section II, Experimental Methods, Section III, Anomalous Hall effect, and Section IV, Experimental Supporting Data.

**References**

1. F. Hellman, A. Hoffmann, Y. Tserkovnyak, G. S. D. Beach, E. E. Fullerton, C. Leighton, A. H. MacDonald, D. C. Ralph, D. A. Arena, H. A. Dürr, P. Fischer, J. Grollier, J. P. Heremans, T. Jungwirth, A. v. Kimel, B. Koopmans, I. N. Krivorotov, S. J. May, A. K. Petford-Long, J. M. Rondinelli, N. Samarth, I. K. Schuller, A. N. Slavin, M. D. Stiles, O. Tchernyshyov, A. Thiaville, B. L. Zink, Interface-induced phenomena in magnetism. *Rev Mod Phys*. **89**, 025006 (2017).
2. F. Giustino, J. H. Lee, F. Trier, M. Bibes, S. M. Winter, R. Valentí, Y. W. Son, L. Taillefer, C. Heil, A. I. Figueroa, B. Plaçais, Q. S. Wu, O. v. Yazyev, E. P. A. M. Bakkers, J. Nygård, P. Forn-Díaz, S. de Franceschi, J. W. McIver, L. E. F. Foa Torres, T. Low, A. Kumar, R. Galceran, S. O. Valenzuela, M. v. Costache, A. Manchon, E. A. Kim, G. R. Schleder, A. Fazzio, S. Roche, The 2021 quantum materials roadmap. *Journal of Physics: Materials*. **3**, 042006 (2021).
3. A. Hirohata, K. Yamada, Y. Nakatani, I.-L. Prejbeanu, B. Diény, P. Pirro, B. Hillebrands, Review on spintronics: Principles and device applications. *J Magn Magn Mater*. **509**, 166711 (2020).

4. F. Matsukura, Y. Tokura, H. Ohno, Control of magnetism by electric fields. *Nat Nanotechnol*. **10**, 209–220 (2015).
5. X. Chen, Q. Wu, L. Zhang, Y. Hao, M.-G. Han, Y. Zhu, X. Hong, Anomalous Hall effect and perpendicular magnetic anisotropy in ultrathin ferrimagnetic $NiCo_2O_4$ films. *Appl Phys Lett*. **120**, 242401 (2022).
6. J. Xia, W. Siemons, G. Koster, M. R. Beasley, A. Kapitulnik, Critical thickness for itinerant ferromagnetism in ultrathin films of SrRuO3. *Phys Rev B*. **79**, 140407 (2009).
7. A. Rastogi, M. Brahlek, J. M. Ok, Z. Liao, C. Sohn, S. Feldman, H. N. Lee, Metal-insulator transition in (111) SrRuO3 ultrathin films. *APL Mater*. **7** (2019), doi:10.1063/1.5109374.
8. D. Du, S. Manzo, C. Zhang, V. Saraswat, K. T. Genser, K. M. Rabe, P. M. Voyles, M. S. Arnold, J. K. Kawasaki, Epitaxy, exfoliation, and strain-induced magnetism in rippled Heusler membranes. *Nat Commun*. **12**, 2494 (2021).
9. A. Bhattacharya, S. J. May, Magnetic Oxide Heterostructures. *Annu Rev Mater Res*. **44**, 65–90 (2014).
10. J. A. Moyer, C. Eaton, R. Engel-Herbert, Highly Conductive SrVO3 as a bottom electrode for functional perovskite oxides. *Adv Mater*. **25**, 3578–82 (2013).
11. L. Zhang, Y. Zhou, L. Guo, W. Zhao, A. Barnes, H.-T. Zhang, C. Eaton, Y. Zheng, M. Brahlek, H. F. Haneef, N. J. Podraza, M. H. W. Chan, V. Gopalan, K. M. Rabe, R. Engel-Herbert, Correlated metals as transparent conductors. *Nat Mater*. **15**, 204–210 (2016).
12. A. P. Mackenzie, The properties of ultrapure delafossite metals. *Reports on Progress in Physics*. **80**, 032501 (2017).
13. T. Harada, Y. Okada, Metallic delafossite thin films for unique device applications. *APL Mater*. **10**, 070902 (2022).
14. H. Takatsu, J. J. Ishikawa, S. Yonezawa, H. Yoshino, T. Shishidou, T. Oguchi, K. Murata, Y. Maeno, Extremely Large Magnetoresistance in the Nonmagnetic Metal PdCoO2. *Phys Rev Lett*. **111**, 056601 (2013).
15. P. J. W. Moll, P. Kushwaha, N. Nandi, B. Schmidt, A. P. Mackenzie, Evidence for hydrodynamic electron flow in $PdCoO_2$. *Science*. **351**, 1061–4 (2016).
16. M. D. Bachmann, A. L. Sharpe, G. Baker, A. W. Barnard, C. Putzke, T. Scaffidi, N. Nandi, P. H. McGuinness, E. Zhakina, M. Moravec, S. Khim, M. König, D. Goldhaber-Gordon, D. A. Bonn, A. P. Mackenzie, P. J. W. Moll, Directional ballistic transport in the two-dimensional metal PdCoO2. *Nature Physics 2022 18:7*. **18**, 819–824 (2022).
17. C. Putzke, M. D. Bachmann, P. McGuinness, E. Zhakina, V. Sunko, M. Konczykowski, T. Oka, R. Moessner, A. Stern, M. König, S. Khim, A. P. Mackenzie, P. J. W. Moll, h/e oscillations in interlayer transport of delafossites. *Science (1979)*. **368**, 1234–1238 (2020).
18. T. Harada, A. Tsukazaki, Control of Schottky barrier height in metal/β-Ga2O3 junctions by insertion of PdCoO2 layers. *APL Mater*. **8**, 041109 (2020).
19. G. Rimal, C. Schmidt, H. Hijazi, L. C. Feldman, Y. Liu, E. Skoropata, J. Lapano, M. Brahlek, D. Mukherjee, R. R. Unocic, M. F. Chisholm, Y. Sun, H. Yu, S. Ramanathan, C.-J. Sun, H. Zhou, S. Oh, Effective reduction of PdCoO2 thin films via hydrogenation and sign tunable anomalous Hall effect. *Phys Rev Mater*. **5**, L052001 (2021).
20. Y. Liu, G. Rimal, P. Narasimhan, S. Oh, Anomalous Hall effect in electrolytically reduced PdCoO2 thin films. *Thin Solid Films*. **751**, 139197 (2022).
21. F. Podjaski, D. Weber, S. Zhang, L. Diehl, R. Eger, V. Duppel, E. Alarcón-Lladó, G. Richter, F. Haase, A. Fontcuberta i Morral, C. Scheu, B. v. Lotsch, Rational strain engineering in delafossite oxides for highly efficient hydrogen evolution catalysis in acidic media. *Nat Catal*. **3**, 55–63 (2020).
22. V. Sunko, P. H. Mcguinness, C. S. Chang, E. Zhakina, S. Khim, C. E. Dreyer, M. Konczykowski, H. Borrmann, P. J. W. Moll, M. König, D. A. Muller, A. P. Mackenzie, Controlled Introduction of Defects to Delafossite Metals by Electron Irradiation. *Phys Rev X*. **10**, 021018 (2020).
23. Q. Lu, H. Martins, J. M. Kahk, G. Rimal, S. Oh, I. Vishik, M. Brahlek, W. C. Chueh, J. Lischner, S. Nemsak, Layer-resolved many-electron interactions in delafossite PdCoO2 from standing-wave photoemission spectroscopy. *Communications Physics 2021 4:1*. **4**, 1–8 (2021).
24. K. P. Ong, J. Zhang, J. S. Tse, P. Wu, Origin of anisotropy and metallic behavior in delafossite PdCoO2. *Phys Rev B*. **81**, 115120 (2010).
25. M. Mekata, T. Sugino, A. Oohara, Y. Oohara, H. Yoshizawa, Magnetic structure of antiferromagnetic PdCrO2 possible degenerate helices on a rhombohedral lattice. *Physica B: Physics of Condensed Matter*. **213–214**, 221–223 (1995).
26. J. B. Goodenough, An interpretation of the magnetic properties of the perovskite-type mixed crystals La1−xSrxCoO3−λ. *Journal of Physics and Chemistry of Solids*. **6**, 287–297 (1958).

27. J. Walter, S. Hara, M. Suzuki, L. S. Suzuki, "Magnetism in Palladium Experimental Results in View of Theoretic Predictions" in *Molecular Low Dimensional and Nanostructured Materials for Advanced Applications*, A. Graja, B. R. Bułka, F. Kajzar, Eds. (Springer Science+Business Media Dordrecht, 2002), vol. 59, pp. 329–333.
28. S. Bouarab, C. Demangeat, A. Mokrani, H. Dreyssé, Onset of magnetism in palladium slabs. *Phys Lett A*. **151**, 103–105 (1990).
29. M. R. Fitzsimmons, S. D. Bader, J. A. Borchers, G. P. Felcher, J. K. Furdyna, A. Hoffmann, J. B. Kortright, I. K. Schuller, T. C. Schulthess, S. K. Sinha, M. F. Toney, D. Weller, S. Wolf, Neutron scattering studies of nanomagnetism and artificially structured materials. *J Magn Magn Mater*. **271**, 103–146 (2004).
30. Q. Song, J. Sun, C. T. Parzyck, L. Miao, Q. Xu, F. V. E. Hensling, M. R. Barone, C. Hu, J. Kim, B. D. Faeth, H. Paik, P. D. C. King, K. M. Shen, D. G. Schlom, Growth of PdCoO2 films with controlled termination by molecular-beam epitaxy and determination of their electronic structure by angle-resolved photoemission spectroscopy. *APL Mater*. **10**, 091113 (2022).
31. F. Mazzola, V. Sunko, S. Khim, H. Rosner, P. Kushwaha, O. J. Clark, L. Bawden, I. Marković, T. K. Kim, M. Hoesch, A. P. Mackenzie, P. D. C. King, Itinerant ferromagnetism of the Pd-terminated polar surface of PdCoO2. *Proc Natl Acad Sci U S A*. **115**, 12956–12960 (2018).
32. F. Mazzola, C. M. Yim, V. Sunko, S. Khim, P. Kushwaha, O. J. Clark, L. Bawden, I. Marković, D. Chakraborti, T. K. Kim, M. Hoesch, A. P. Mackenzie, P. Wahl, P. D. C. King, Tuneable electron–magnon coupling of ferromagnetic surface states in PdCoO2. *NPJ Quantum Mater*. **7**, 1–6 (2022).
33. T. Harada, K. Sugawara, K. Fujiwara, M. Kitamura, S. Ito, T. Nojima, K. Horiba, H. Kumigashira, T. Takahashi, T. Sato, A. Tsukazaki, Anomalous Hall effect at the spontaneously electron-doped polar surface of PdCoO2 ultrathin films. *Phys Rev Res*. **2**, 013282 (2020).
34. H. Guo, S. Dong, P. Rack, J. Budai, C. Beekman, Z. Gai, W. Siemons, C. Gonzalez, R. Timilsina, A. T. Wong, A. Herklotz, P. C. Snijders, E. Dagotto, T. Z. Ward, Strain Doping: Reversible Single-Axis Control of a Complex Oxide Lattice via Helium Implantation. *Phys Rev Lett*. **114**, 256801 (2015).
35. E. Skoropata, A. R. Mazza, A. Herklotz, J. M. Ok, G. Eres, M. Brahlek, T. R. Charlton, H. N. Lee, T. Z. Ward, Post-synthesis control of Berry phase driven magnetotransport in SrRuO3 films. *Phys Rev B*. **103**, 085121 (2021).
36. A. Herklotz, A. T. Wong, T. Meyer, M. D. Biegalski, H. N. Lee, T. Z. Ward, Controlling Octahedral Rotations in a Perovskite via Strain Doping. *Sci Rep*. **6**, 26491 (2016).
37. A. R. Mazza, A. Miettinen, Z. Gai, X. He, T. R. Charlton, T. Z. Ward, M. Conrad, G. Bian, E. H. Conrad, P. F. Miceli, The structural modification and magnetism of many-layer epitaxial graphene implanted with low-energy light ions. *Carbon N Y*. **192**, 462–472 (2022).
38. P. Giannozzi, S. Baroni, N. Bonini, M. Calandra, R. Car, C. Cavazzoni, D. Ceresoli, G. L. Chiarotti, M. Cococcioni, I. Dabo, A. D. Corso, S. de Gironcoli, S. Fabris, G. Fratesi, R. Gebauer, U. Gerstmann, C. Gougoussis, A. Kokalj, M. Lazzeri, L. Martin-Samos, N. Marzari, F. Mauri, R. Mazzarello, S. Paolini, A. Pasquarello, L. Paulatto, C. Sbraccia, S. Scandolo, G. Sclauzero, A. P. Seitsonen, A. Smogunov, P. Umari, R. M. Wentzcovitch, QUANTUM ESPRESSO: a modular and open-source software project for quantum simulations of materials. *Journal of Physics: Condensed Matter*. **21**, 395502 (2009).
39. J. Kim, A. D. Baczewski, T. D. Beaudet, A. Benali, M. C. Bennett, M. A. Berrill, N. S. Blunt, E. J. L. Borda, M. Casula, D. M. Ceperley, S. Chiesa, B. K. Clark, R. C. Clay, K. T. Delaney, M. Dewing, K. P. Esler, H. Hao, O. Heinonen, P. R. C. Kent, J. T. Krogel, I. Kylänpää, Y. W. Li, M. G. Lopez, Y. Luo, F. D. Malone, R. M. Martin, A. Mathuriya, J. McMinis, C. A. Melton, L. Mitas, M. A. Morales, E. Neuscamman, W. D. Parker, S. D. Pineda Flores, N. A. Romero, B. M. Rubenstein, J. A. R. Shea, H. Shin, L. Shulenburger, A. F. Tillack, J. P. Townsend, N. M. Tubman, B. van der Goetz, J. E. Vincent, D. C. Yang, Y. Yang, S. Zhang, L. Zhao, QMCPACK : an open source ab initio quantum Monte Carlo package for the electronic structure of atoms, molecules and solids. *Journal of Physics: Condensed Matter*. **30**, 195901 (2018).
40. P. R. C. Kent, A. Annaberdiyev, A. Benali, M. C. Bennett, E. J. Landinez Borda, P. Doak, H. Hao, K. D. Jordan, J. T. Krogel, I. Kylänpää, J. Lee, Y. Luo, F. D. Malone, C. A. Melton, L. Mitas, M. A. Morales, E. Neuscamman, F. A. Reboredo, B. Rubenstein, K. Saritas, S. Upadhyay, G. Wang, S. Zhang, L. Zhao, QMCPACK: Advances in the development, efficiency, and application of auxiliary field and real-space variational and diffusion quantum Monte Carlo. *J Chem Phys*. **152**, 174105 (2020).
41. W. M. C. Foulkes, L. Mitas, R. J. Needs, G. Rajagopal, Quantum Monte Carlo simulations of solids. *Rev Mod Phys*. **73**, 33–83 (2001).

42. J. M. Ok, M. Brahlek, W. S. Choi, K. M. Roccapriore, M. F. Chisholm, S. Kim, C. Sohn, E. Skoropata, S. Yoon, J. S. Kim, H. N. Lee, Pulsed-laser epitaxy of metallic delafossite PdCrO2 films. *APL Mater*. **8**, 051104 (2020).
43. N. Wolff, T. Schwaigert, D. Siche, D. G. Schlom, D. Klimm, Growth of CuFeO2 single crystals by the optical floating-zone technique. *J Cryst Growth*. **532**, 125426 (2020).
44. M. Brahlek, G. Rimal, J. M. Ok, D. Mukherjee, A. R. Mazza, Q. Lu, H. N. Lee, T. Z. Ward, R. R. Unocic, G. Eres, S. Oh, Growth of metallic delafossite PdCoO2 by molecular beam epitaxy. *Phys Rev Mater*. **3**, 093401 (2019).
45. A. B. Mei, I. Gray, Y. Tang, J. Schubert, D. Werder, J. Bartell, D. C. Ralph, G. D. Fuchs, D. G. Schlom, Local Photothermal Control of Phase Transitions for On-Demand Room-Temperature Rewritable Magnetic Patterning. *Advanced Materials*. **32**, 2001080 (2020).
46. J. Fassbender, J. McCord, Magnetic patterning by means of ion irradiation and implantation. *J Magn Magn Mater*. **320**, 579–596 (2008).
47. S. Kim, S. Lee, J. Ko, J. Son, M. Kim, S. Kang, J. Hong, Nanoscale patterning of complex magnetic nanostructures by reduction with low-energy protons. *Nat Nanotechnol*. **7**, 567–571 (2012).
48. E. Morenzoni, H. Glückler, T. Prokscha, H. P. Weber, E. M. Forgan, T. J. Jackson, H. Luetkens, C. Niedermayer, M. Pleines, M. Birke, A. Hofer, J. Litterst, T. Riseman, G. Schatz, Low-energy μSR at PSI: present and future. *Physica B Condens Matter*. **289–290**, 653–657 (2000).
49. T. Prokscha, E. Morenzoni, K. Deiters, F. Foroughi, D. George, R. Kobler, A. Suter, V. Vrankovic, The new beam at PSI: A hybrid-type large acceptance channel for the generation of a high intensity surface-muon beam. *Nucl Instrum Methods Phys Res A*. **595**, 317–331 (2008).
50. E. Morenzoni, H. Glückler, T. Prokscha, R. Khasanov, H. Luetkens, M. Birke, E. M. Forgan, Ch. Niedermayer, M. Pleines, Implantation studies of keV positive muons in thin metallic layers. *Nucl Instrum Methods Phys Res B*. **192**, 254–266 (2002).
51. W. Eckstein, *Computer Simulation of Ion-Solid Interactions* (Springer Berlin Heidelberg, Berlin, Heidelberg, 1991), vol. 10.
52. J. F. Ziegler, M. D. Ziegler, J. P. Biersack, SRIM – The stopping and range of ions in matter (2010). *Nucl Instrum Methods Phys Res B*. **268**, 1818–1823 (2010).
53. R. Morrow, K. Samanta, T. Saha Dasgupta, J. Xiong, J. W. Freeland, D. Haskel, P. M. Woodward, Magnetism in Ca2CoOsO6 and Ca2NiOsO6: Unraveling the Mystery of Superexchange Interactions between 3d and 5d Ions. *Chemistry of Materials*. **28**, 3666–3675 (2016).
54. M. W. Haverkort, Z. Hu, J. C. Cezar, T. Burnus, H. Hartmann, M. Reuther, C. Zobel, T. Lorenz, A. Tanaka, N. B. Brookes, H. H. Hsieh, H.-J. Lin, C. T. Chen, L. H. Tjeng, Spin State Transition in LaCoO3 Studied Using Soft X-ray Absorption Spectroscopy and Magnetic Circular Dichroism. *Phys Rev Lett*. **97**, 176405 (2006).
55. J. Crangle, W. R. Scott, Dilute Ferromagnetic Alloys. *J Appl Phys*. **36**, 921 (2004).
56. N. Nagaosa, J. Sinova, S. Onoda, A. H. MacDonald, N. P. Ong, Anomalous Hall effect. *Rev Mod Phys*. **82**, 1539–1592 (2010).
57. S. Ikeda, K. Miura, H. Yamamoto, K. Mizunuma, H. D. Gan, M. Endo, S. Kanai, J. Hayakawa, F. Matsukura, H. Ohno, A perpendicular-anisotropy CoFeB–MgO magnetic tunnel junction. *Nat Mater*. **9**, 721–724 (2010).
58. Y. Fujishiro, N. Kanazawa, R. Kurihara, H. Ishizuka, T. Hori, F. S. Yasin, X. Yu, A. Tsukazaki, M. Ichikawa, M. Kawasaki, N. Nagaosa, M. Tokunaga, Y. Tokura, Giant anomalous Hall effect from spin-chirality scattering in a chiral magnet. *Nat Commun*. **12**, 317 (2021).
59. S.-Y. Yang, Y. Wang, B. R. Ortiz, D. Liu, J. Gayles, E. Derunova, R. Gonzalez-Hernandez, L. Šmejkal, Y. Chen, S. S. P. Parkin, S. D. Wilson, E. S. Toberer, T. McQueen, M. N. Ali, Giant, unconventional anomalous Hall effect in the metallic frustrated magnet candidate, KV3Sb5. *Sci Adv*. **6** (2020), doi:10.1126/sciadv.abb6003.
60. K. F. Mak, J. Shan, D. C. Ralph, Probing and controlling magnetic states in 2D layered magnetic materials. *Nature Reviews Physics*. **1**, 646–661 (2019).
61. R. Bali, S. Wintz, F. Meutzner, R. Hübner, R. Boucher, A. A. Ünal, S. Valencia, A. Neudert, K. Potzger, J. Bauch, F. Kronast, S. Facsko, J. Lindner, J. Fassbender, Printing Nearly-Discrete Magnetic Patterns Using Chemical Disorder Induced Ferromagnetism. *Nano Lett*. **14**, 435–441 (2014).

**Acknowledgments**

Processing, electronic characterization and manuscript were supported by the U.S. Department of Energy (DOE), Office of Science, Basic Energy Sciences (BES), Materials Sciences and Engineering Division.

Spectroscopy was supported by the U.S. Department of Energy, Office of Science, National Quantum Information Science Research Centers, Quantum Science Center.

Part of the theory work (quantum Monte Carlo and density functional theory using Quantum ESPRESSO) has been supported by the U.S. Department of Energy, Office of Science, Basic Energy Sciences, Materials Sciences and Engineering Division, as part of the Computational Materials Sciences Program and Center for Predictive Simulation of Functional Materials. Initial theory screening (density functional theory using VASP) that formed the basis of current theory-work was supported by the Center for Nanophase Materials Sciences (CNMS), which is a U.S. Department of Energy, Office of Science User Facility at Oak Ridge National Laboratory. An award of computer time was provided by the Innovative and Novel Computational Impact on Theory and Experiment (INCITE) program. Part of this research (quantum Monte Carlo) used resources of the Oak Ridge Leadership Computing Facility, which is a DOE Office of Science User Facility supported under Contract No. DE-AC05-00OR22725. Part of this research (quantum Monte Carlo) used resources of the Argonne Leadership Computing Facility, which is a DOE Office of Science User Facility supported under Contract No. DE-AC02-06CH11357. Part of this research (density functional theory) used resources of the National Energy Research Scientific Computing Center (NERSC), a U.S. Department of Energy Office of Science User Facility located at Lawrence Berkeley National Laboratory, operated under Contract No. DE-AC02-05CH11231. This paper describes objective technical results and analysis. Any subjective views or opinions that might be expressed in the paper do not necessarily represent the views of the U.S. Department of Energy or the United States Government.

The use of the Advanced Photon Source, Argonne National Laboratory was supported by the U.S. Department of Energy, Office of Science, Basic Energy Sciences, under Contract No. DE-AC02-06CH11357.

The use of the Swiss Muon Source (SµS) and Swiss Light Source (SLS), Paul Scherrer Institute, Villigen, Switzerland. Muon spectroscopy was partially supported by Laboratory Directed Research and Development Program of Oak Ridge National Laboratory, managed by UT-Battelle, LLC, for the U. S. Department of Energy.

Synthesis was supported by National Science Foundation (NSF) Grant No. DMR2004125 and Army Research Office (ARO) Grant No. W911NF2010108.

Work at CINT, an Office of Science User Facility operated for the U.S. Department of Energy Office of Science through the Los Alamos National Laboratory. Los Alamos National Laboratory is operated by Triad National Security, LLC, for the National Nuclear Security Administration of U.S. Department of Energy (Contract No. 89233218CNA000001).

## Author contributions

M.B., A.R.M, and T.Z.W conceived the project;
A.A., A.P., J.T.K., P.G. performed the theoretical calculations, as directed by P.G.;
M.B., A.R.M., and M.C. performed the implantation in direction of T.Z.W.;
G.R., and S.O. performed the synthesis in coordination with M.B;
M.B., M.C., and G.E. performed the structural characterization;
J.W.F. performed the XMCD;
M.B. performed the transport measurements with help from J.L., and Y.-Y. P., with input from G.H.;
R.G.M. performed the spectroscopy with assistance from Y.C., and J.M.;
Z.S., M.B., T.Z.W., and J.S.G. performed muon spin rotations with assistance from T.P. and A.S.;
M.B. wrote the manuscript with input from all coauthors and directed the work.

## Competing interests
There are no competing interests.

## Data and materials availability

The data supporting this study's findings are available from the corresponding author upon reasonable request.

# Supplementary Information for

## Emergent magnetism with continuous control in the ultrahigh conductivity layered oxide PdCoO$_2$

Matthew Brahlek[1*], Alessandro R. Mazza[1,2], Abdulgani Annaberdiyev[3], Michael Chilcote[1], Gaurab Rimal[4], Gábor B. Halász[1], Anh Pham[3], Yun-Yi Pai[1], Jaron T. Krogel[1], Jason Lapano[1], Benjamin J. Lawrie[1], Gyula Eres[1], Jessica McChesney[5], Thomas Prokscha[6], Andreas Suter[6], Seongshik Oh[4], John W. Freeland[5], Yue Cao[7], Jason S. Gardner[1], Zaher Salman[6], Robert G. Moore[1], Panchapakesan Ganesh[3#], T. Zac Ward[1]

[1]Materials Science and Technology Division, Oak Ridge National Laboratory, Oak Ridge, TN, 37831, USA
[2]Center for Integrated Nanotechnologies, Los Alamos National Laboratory, Los Alamos, NM
[3]Center for Nanophase Materials Sciences, Oak Ridge National Laboratory, Oak Ridge, TN, 37831, USA
[4]Department of Physics and Astronomy, Rutgers, The State University of New Jersey, Piscataway, NJ, 08854, USA
[5]Advanced Photon Source, Argonne National Laboratory, Lemont, IL, 60439, USA
[6]Laboratory for Muon Spin Spectroscopy, Paul Scherrer Institute, CH-5232 Villigen PSI, Switzerland
[7]Materials Science Division, Argonne National Laboratory, Lemont, IL, 60439, USA
Correspondence should be addressed to *brahlekm@ornl.gov, #ganeshp@ornl.gov

Supplementary Information includes: Section I, Ab-initio DFT and QMC Supporting Data, Section II, Experimental Methods, and Section III, Experimental Supporting Data.

## I. Ab-initio DFT and QMC Supporting Data

### A. Methods

PdCoO$_2$ shows both itinerant and localized electronic behavior. Prior efforts to use density functional theory (DFT) to describe PdCoO$_2$ have invoked the Hubbard U correction (DFT+U) to describe strong Coulomb interactions between the electrons in the Co 3d band. The choice of U value modifies the level of electron localization and spin polarization in the Co sites and thus should affect the magnetic properties. For instance, a large value of $U = 4$ eV was found necessary in PtCoO$_2$ to account for correlations in CoO$_2$(*1*). Since it is not known apriori what value of $U$ to use for accurate prediction of the magnetic ordering and related magnetic properties, we use quantum Monte Carlo (QMC) based calculations. QMC fully captures the effects of strong- and weak-electron correlations (with some approximations) and is arguably the most accurate electronic-structure method to predict the ground state properties of a solid. We use the fixed-node/fixed-phase diffusion Monte Carlo (DMC) flavor of QMC, where the walkers are propagated in real space. There are two major approximations in QMC: (i) the nodal surface of the wavefunction is fixed to that of the trial wave function, and (ii) pseudopotentials are used to reduce the total number of electrons explicitly considered while solving the Schrödinger equation. We perform DFT+U calculations using the PBE functional(*2*) to generate the trial wave functions. Due to the variational theorem, the total energy obtained from DMC can be variationally optimized to predict the optimal Hubbard U parameter as discussed later.

We used the correlation consistent effective core potentials (ccECP) to accurately represent the valence electrons in DFT+U and Fixed-Node Diffusion Monte Carlo (FNDMC) calculations. A high 650 Ry kinetic energy cutoff was used in plane-wave DFT calculations performed using the Quantum ESPRESSO package. A uniform [28×28×16] (*x*×*y*×*z*) **k**-mesh was used to sample the Brillouin zone unless otherwise specified. FNDMC, performed using the QMCPACK package, had a timestep of 0.01 Ha$^{-1}$ and T-moves localization algorithm(*3*) was used in all cases. Trial wave functions were single-reference, Slater-Jastrow type functions where the Jastrow terms included one-body, two-body, and three-body terms. The experimental structure, shown below in Fig. S1, was used for all calculations.

```
CRYSTAL
 PRIMVEC
   2.82999992    0.00000000    0.00000000
  -1.41499996    0.81695061    5.91433334
   1.41499996   -2.45085183    0.00000000
 PRIMCOORD
 4 1
   46   0.00000000    0.00000000    0.00000000
   27   1.41499996   -0.81695061    2.95716667
    8   0.00000000    0.00000000    1.97302116
    8   2.82999992   -1.63390122    3.94131218
```

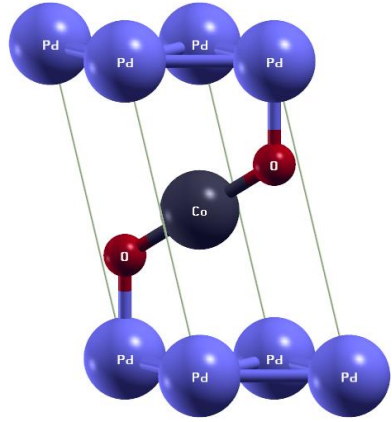

**Fig. S1** The employed experimental structure listed in XSF format (all coordinates are in Angstrom units) and the panel on the right shows the primitive cell.

### B. FNDMC Total Energies

In Fig. S2, we plot the FNDMC total energies for a [2×2×2] supercell of $PdCoO_2$ for various trial wave functions obtained from the DFT+U method. Hubbard U is applied to the Co atoms, which are more localized than Pd atoms. Since the FNDMC method is variational, it can probe for the optimal trial wave function and find the appropriate DFT functional for a given material and state(4–6). For this material, we find that the $U \approx 1$ eV trial wave functions provide the lowest energies for both nonmagnetic and magnetic states. This suggests that the same appropriate value of Hubbard U can be used for various states when studying the relative characteristics. For instance, the same $U \approx 1$ eV value could be used when studying the response to stress for various magnetic states. Note that determining the optimal value of +U is not trivial in metals since other methods which rely on the calculation of the self-consistent dielectric constant are not applicable(7). Fig. S2 also demonstrates another important effect of applying the Hubbard U on nonmagnetic and magnetic states - although the optimal value +U is the same, the magnetic states display larger energy changes as the Hubbard U value increases (bottom right). The improvement in energy going from FNDMC/PBE to FNDMC/PBE+U (1 eV) is more pronounced in the magnetic state when compared to the nonmagnetic state. This points to the critical role of using an appropriate Hubbard U in systems with significant magnetism.

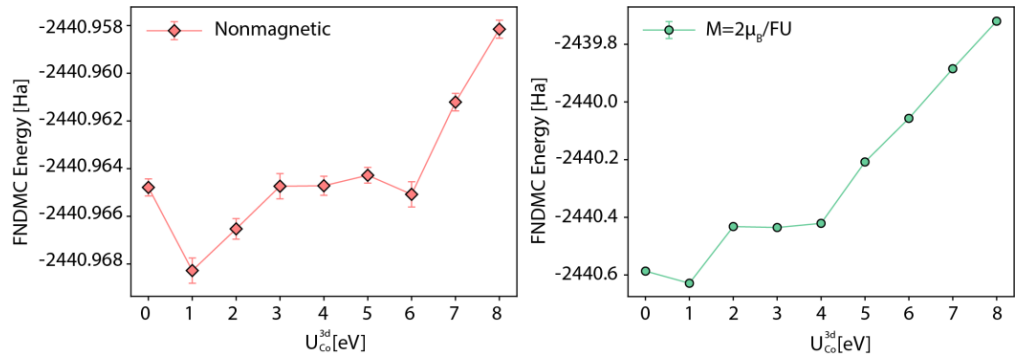

**Fig. S2** FNDMC energies [Ha] for $[2 \times 2 \times 2]$ supercell of $PdCoO_2$ for various trial wave functions. Energies are shown for (left) nonmagnetic state, (right) spin magnetization of $M = 2\mu_B$ per formula unit (FU). Note that the magnetic state display larger energy changes as a function of Hubbard-U on Co-3d states.

### C. DFT+U Energies

In Fig. S3, we plot the total energies of the primitive cell using the DFT+U method as the c-axis, out-of-plane strain is applied. The Hubbard $U_{Co}$ is taken to be 1 eV here and in the rest of the calculations (see the previous section for more details). A few aspects can be extracted from this plot. First, as the system is uniaxially compressed (negative strain), the total energies increase as expected – this indicates the near optimality of the employed experimental

Page 19

structure. Second, near ambient conditions, the nonmagnetic state has much lower energy than the magnetic states (here $M = 2\mu_B$ per formula unit (FU) and $M = 4\mu_B$/FU) – this agrees with the experimentally observed paramagnetic state. Finally, near ~30% strain, the nonmagnetic and magnetic state energies are very close, and magnetic states eventually result in lower energies as the system is strained.

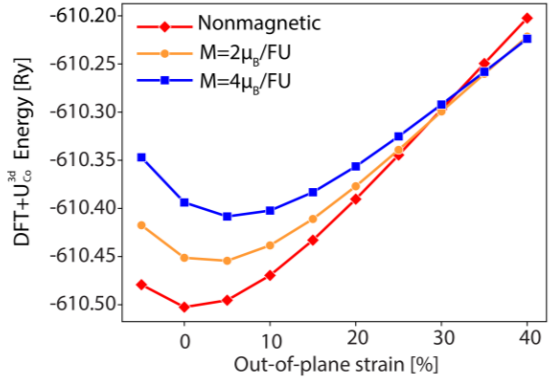

**Fig. S3** Energies of nonmagnetic and magnetic states for out-of-plane strained $PdCoO_2$ using DMC-benchmarked DFT+U. See main text Fig. 1d for energy difference among the magnetic and non-magnetic states.

### D. DFT+U Effective Charges

In Fig. S4, we plot the Löwdin atomic charges using the DFT+U method as the c-axis, out-of-plane strain is applied. Overall, we see a charge transfer from the Pd layer to the $CoO_2$ sites as the crystal is strained. The charge transfer can be seen as a decrease of the charge magnitude on the Pd and O atoms while Co stays relatively flat. In other words, $CoO_2$ sites transfer electrons to the Pd layer. We conjecture that this charge transfer is partly responsible for the emergence of localized moments in Co atoms and itinerant moments on the Pd layer.

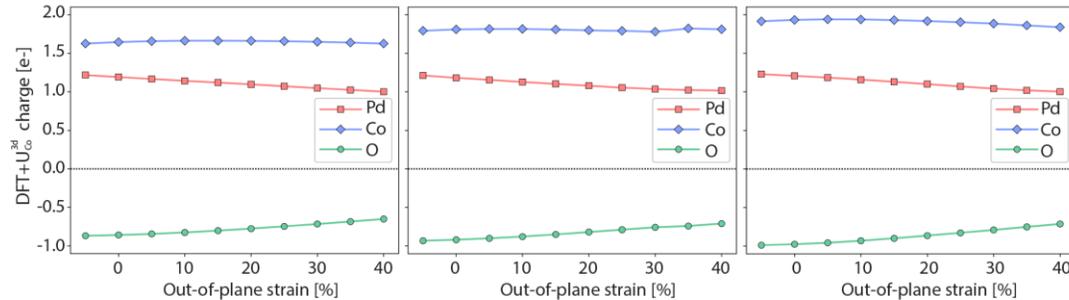

**Fig. S4** Löwdin effective atomic charges of nonmagnetic and magnetic structures as a function of out-of-plane strained $PdCoO_2$ using DMC-benchmarked DFT+U. (Left) nonmagnetic, (middle) $M = 2\mu_B$/FU, (right) $M = 4\mu_B$/FU. Note the steady decrease of Pd and O charges in all cases.

### E. DFT+U Density of States

Fig. S5 shows the atom-specific changes in density of states (DOS) as the crystal is strained for nonmagnetic and magnetic phases. These results give insight into how the electronic states are modified for the magnetic phases and the previously discussed charge transfer. First, we see that the DOS at the Fermi level significantly increases as the system is strained. This is a key ingredient in the simplified Stoner model for the ferromagnetism to occur, namely, $DOS(E_F)U > 1$. More specifically, these calculated results imply that in implanted samples the local distortion will cause a large enhancement of the density of states around the helium atom and thus creates an area with a magnetic moment.

Another result of these calculations that can be directly probed experimentally is that there is a larger Co character at the Fermi level in strained magnetic states when compared to the strained nonmagnetic state. This supports that the magnetism has both an itinerant (from Pd) as well as a localized (from Co) character, as elaborated in the main article.

Interestingly, we find that in both magnetic states, the Pd atoms are spin-polarized. Indeed, even when unstrained, an intermediate-spin configuration results in a significant Pd moment of the unit cell. Since the majority of carriers at the Fermi level are derived from the itinerant Pd states, this should lead to spin-polarized transport signatures as well, as seen in the Hall effect measurements shown in the main text.

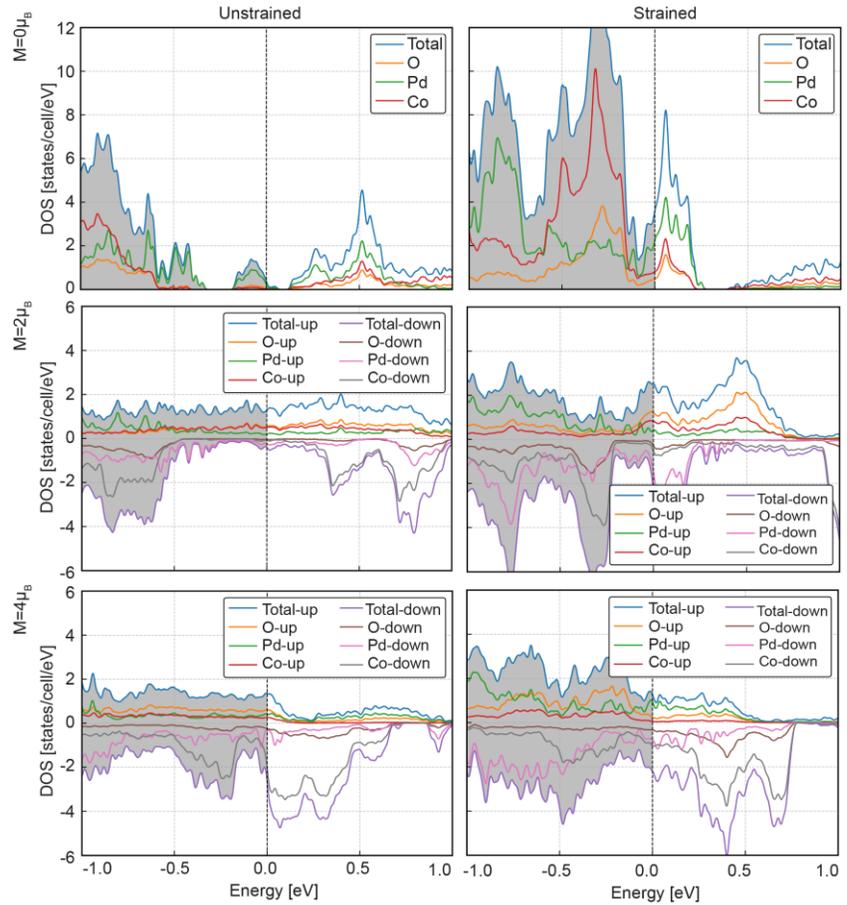

**Fig. S5** Density of states (DOS) of $PdCoO_2$ using DMC-benchmarked DFT+U. Left column corresponds to unstrained, ambient condition structures. Right column is structures with a 25% strain of the out-of-plane axis. Top row is the nonmagnetic state, middle row is $M = 2\mu_B$/FU and bottom row is $M = 4\mu_B$/FU. Positive and negative DOS represent the contributions from the up and down electrons.

### F. VMC Fermi Surface

For the calculation of the VMC Fermi surface (shown in Fig. S6 and compared to experimental results in the main text Fig. 2b), DFT+U SCF calculations were carried out with PBE functional(*2*) and $U = 1$ eV applied on the Co atoms. These calculations used [28×28×16] (*x*×*y*×*z*) k-point grid in the primitive cell, which was dense enough to converge the total energy. The variational Monte Carlo (VMC) calculations of the Fermi surface were based on 32 atom supercells ([2×2×2] of the primitive cell) of bulk $PdCoO_2$ using a [14×14×1] twist grid, which is equivalent to a [28×28×2] Monkhorst-Pack k-point grid in the primitive cell. At each supercell twist, the electron momentum distribution was accumulated and then backfolded into the first Brillouin zone following the Lock-Crisp-West(*8*) procedure giving the distribution of primitive cell electron occupations in k-space. Prior to backfolding, the VMC momentum distribution was augmented with DFT data at large *k* values in the evanescent region, which restores the norm following backfolding. The Fermi surface was obtained directly from the location of the step discontinuity in the distribution of electronic occupations.

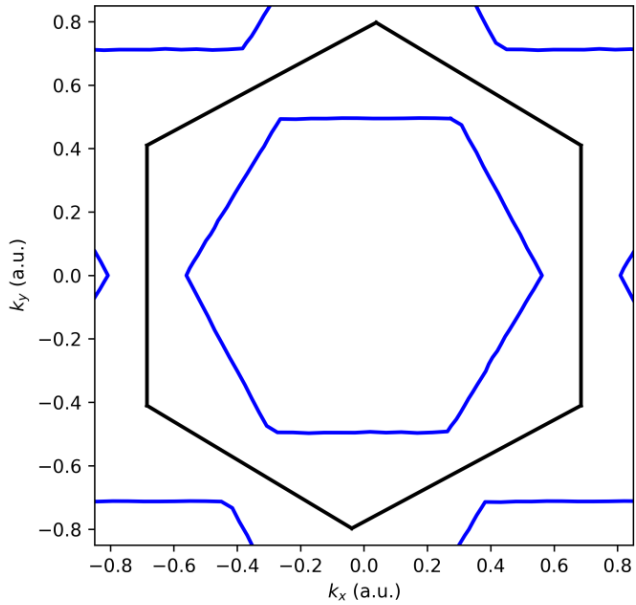

**Fig. S6** Fermi surface calculated using QMC. The blue inner pockets are the bulk Fermi surfaces and the black outer pocket is the Brillouin zone. The units are inverse Bohr (~0.529 Å).

## II. Experimental methods
### A. Synthesis and implantation

The films used in this study were grown using a two-step molecular beam epitaxy growth process, as reported in Ref.(9) . The films were grown on c-axis $Al_2O_3$ substrates that were heated to 700 °C in RF atomic oxygen at a pressure of $4\times10^{-6}$ Torr and a power of 450 Watts then cooled to 300 °C. $PdCoO_2$ was subsequently grown in a layer-by-layer fashion where Pd-Co-Pd-Co- etc layers were alternately grown. To improve crystallinity and overall properties the films were ex-situ annealed to 750 °C in air. This process enables creating thick films with large residual resistance ratios, indicating high crystalline quality.

The helium implantation was done in a home-built ion implantation chamber. The samples were pumped down to a base pressure less than $1\times10^{-7}$ Torr, at which point helium was leaked into the chamber to a pressure of $6\times10^{-5}$ Torr. Once the pressure was stabilized the helium beam was ramped to 5 kV and the flux carefully measured to ensure the correct dosage and stability. The samples were implanted for a calculated time to achieve the desired dose. Post-implantation the samples were characterized structurally and electrically: The structural measurements were done using a 4-circle X-ray diffractometer using copper $k_{\alpha 1}$ radiation.

### B. Transport

Transport measurements were done in a Quantum Design Physical Property Measurement System up to maximum magnetic field of 9 T and base temperature of 2 K with a standard AC bridge. The measurements from 1.5 K down to ~40 mK were performed up to 14 T in an Oxford Triton dilution refrigerator using a lock-in amplifier setup. Contacts were made using pressed indium wires in van der Pauw geometry and a wire bonder for the Hall bar data shown in the main text. Sweeps were performed from the $-H_{max}$ to $+H_{max}$, then swept from $+H_{max}$ to $-H_{max}$. This enable proper symmetrization to remove any small mixing of $R_{xx}$ and $R_{xy}$ that typically occurs in transport measurements. The anomalous Hall resistance was obtained via $R_{xy,A} = R_{xy,m} - L \cdot H$ with $R_{xy,m}$ being the measured Hall response, $L$ linear slope at high field (Lorentz free carrier response), $H$ is the applied magnetic field. The conductivities ($\sigma_{xy}$ and $\sigma_{xx}$) were calculated by inverting the resistivity matrix ($\rho_{xy}$ and $\rho_{xx}$), which yields $\sigma_{xy}=\rho_{xy}/(\rho_{xy}^2+\rho_{xx}^2)$ and $\sigma_{xx}=\rho_{xx}/(\rho_{xy}^2+\rho_{xx}^2)$.

### C. X-ray Photoemission Spectroscopy

Soft x-ray photoelectron spectroscopy (XPS) was carried out at the beamline 29-ID-C of the Advanced Photon Source (APS). The $PdCoO_2$ samples were synthesized ex-situ and loaded into the ultrahigh vacuum chamber. The data were collected at 20 K, with a base pressure better than $5 \times 10^{-10}$ Torr.

### D. Muon Spin Relaxation

Muon spin relaxation was performed at the low-energy muon spin rotation (LE-μSR) beamline at the LEM spectrometer(10, 11) at the Paul Scherrer Institute using positively charged muons $\mu^+$. The muons impinge one by one onto the sample with 100% spin-polarization oriented normal to the momentum direction. Implantation energies from 2-13 keV were used and a transverse magnetic field of 100 G orientated normal to the sample surface. The emitted positrons from the subsequent muon decay were sensed using an array of 8 detectors arranged around the axis of the muon beam, as shown in Fig. S7a. The detectors were arranged in two ring-arrays, upstream and downstream of the sample, each containing 4 detectors that were located at positions top, bottom, left and right. Since we were only interested in resolving spin rotation in the sample plane, each detector in the same axial position upstream and downstream were summed. These detectors therefore measured the time histogram of the decay of the muons during precession.

The time dependence of the positrons emitted from the muon decay was individually analyzed using:

(1) $$N_{e+}(t) = N_0 \left(1 + A_\mu P(T)\right) e^{-t/\tau_\mu}$$

Here $N_{e+}(t)$ is the signal of each position detector, $N_0$ is the number of positrons measured at $t = 0$, $A_\mu$ is the initial asymmetry parameter, $P(t)$ is the time-dependent spin polarization, and $\tau_\mu$ is the lifetime of the muon, 2.2 μs. With an applied transverse magnetic field $P(T)$ is a damped oscillatory function which captures the precession of the muon in the applied and internal magnetic field, which is given by

(2) $$A_\mu P(T) = A_\mu e^{-(\sigma t)^2/2} \cos\left(\omega_\mu t + \phi\right).$$

Here, the cosine term captures the precession of the muon with Larmor frequency $\omega_\mu = \gamma_\mu B$, $\gamma_\mu$ is the gyromagnetic ratio of the muon, $\gamma_\mu = 0.8516$ Mrad/mT, and $B$ is the average magnetic field sensed by the muon, composed of applied plus internal fields. ϕ is the phase factor that captures the 90° offset of the detectors. The Gaussian damping term captures the distribution width of the magnetic field via the parameter σ. Here, the key experimental data is encoded within the parameters $B$, $\sigma$ and $A_\mu$, which was extracted by fitting the experimental data for the detector arrays to equations 1 and 2 using the musrfit package(*12*). In a material which undergoes a magnetic transition, muons landing in magnetic regions of the sample experience a large magnetic field and depolarize almost immediately due to fast precession. However, muons landing in paramagnetic regions experience predominantly the externally applied field and precess at a frequency close to the Larmor frequency. Therefore, the measured initial asymmetry $A_\mu$, is proportional to the non-magnetic volume fraction. The volume fraction is given by $(A_\mu(300K) - A_\mu(T))/(A_\mu(300K) - A_{\mu,0})\times 100\%$, where $A_{\mu,0}$ is the instrument's background ~0.03 (*10*, *11*). As shown in the main text we can accurately calculate the magnetic volume fraction from the initial asymmetry versus temperature, which captures the onset and ordering of $PdCoO_2$ in a ferromagnetic phase.

To get an estimation for the stopping depth distribution of the muons within the films we simulated the stopping profile using TRIM.SP(*13*, *14*). Figure 2d shows the resulting implantation profile for $PdCoO_2$ assuming a density of 7.94 gcm$^{-3}$. This shows that for a kinetic energy of 2 keV, the muons should stop within 10 - 15 nm of the surface of $PdCoO_2$. The energy dependence of the asymmetry is shown in Fig. S7b taken at room temperature. We attribute the drop in asymmetry to muons stopping in the sapphire substrate where they form muonium (i.e. a bound state of a muon and an electron). Muons that form a muonium state lose there polarization due to the string hyperfine interaction with the electron. From these measurements we note that the muons reach the substrate at implantation energy larger than 7 keV. This is slightly lower than the 8 keV energy needed to reach the substrate from TRIM.SP simulations (see Fig. 2d, which shows that the depth is likely a slight underestimate i.e. the muons penetrate deeper than the simulation, possibly due to slightly underestimated density or film thickness).

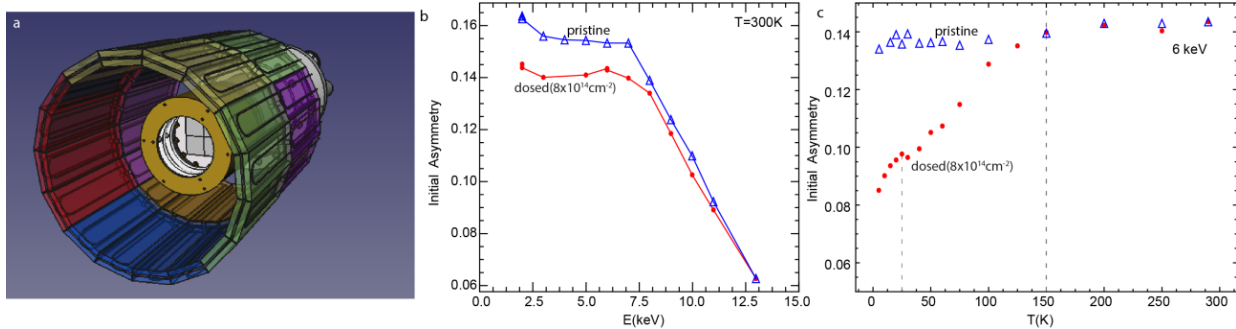

**Fig. S7 a**, Schematic of the detector array (outer ring) used at the LEM spectrometer(*10*, *11*) at the Paul Scherrer Institute. Each colored panel is one of eight individual detectors. The momentum of the low-energy muons is along the axis of the detector array and their spin oriented orthogonal to the axis. They impinge on the samples (small squares) at the center of the array. **b-c**, Energy (b) and temperature (c) dependence of the asymmetry factor for pristine and implanted ($8\times 10^{14}$ cm$^{-2}$) films. The data in (b) was taken at room temperature (b) and the data in (c) was taken at fixed energy, both in a transverse magnetic field of 100 G. The vertical lines in (c) correspond to the ordering temperatures discussed in the main text.

### III.   Anomalous Hall effect
#### A. Background Information

The Hall effect is cause by the buildup of charge in the direction transverse to an applied electric field. Semiclassically, this is due to the deflection of charged carriers moving in response to an applied current. This is simply due to the charges in a material executing cyclotron orbits due to the Lorentz force (the ordinary Hall effect). In a magnetic material the moving carriers respond both to an applied magnetic field but also the internal magnetization. This results in the anomalous Hall effect. Although known for a long time, a full understanding of this effect has only recently been settled which took a full description of the quantum mechanical effects in a solid. A complete account can be found in Ref. (*15*).

A basic understanding of the anomalous Hall effect which is needed to understand the present data in helium implanted $PdCoO_2$ requires understanding the various origins of the transverse deflection of the applied current. Although experimentally the resistivity is the quantity that is directly measured (simply the ratio of the applied current relative to the net voltage drop), theoretically the conductivity is a more natural and an easily handled quantity. Experimentally, the conductivity can be derived by inverting the resistivity matrix and taking the longitudinal ($\sigma_{xx} =$ $\rho_{xx}/(\rho_{xy}^2 + \rho_{xx}^2))$ and transverse ($\sigma_{xy} = \rho_{xy}/(\rho_{xy}^2 + \rho_{xx}^2)$) components. As such, theory finds there are several regimes where the anomalous Hall conductivity is dominated by certain physical mechanisms, which gives a framework to understand our observations for the magnetic turn on in high-conductivity $PdCoO_2$.

First, in the highest conductivity regime ($\sigma_{xx} > 1 M\Omega^{-1}\cdot cm^{-1}$ or $\rho_{xx} < 1 \mu\Omega\cdot cm$), extrinsic effects are the dominant cause of the transverse current. This is due to magnetic sites favoring electrons being scattered in one direction more so than the opposite. In turn, this asymmetry causes a buildup of charge on a specific side of the material. This has been attributed, for example, to skew scattering where the momentum direction of the carriers is changed based on the direction of the local magnetic site. As such, by changing the moment direction via an applied magnetic field the direction the carriers are deflected also reverses. Therefore, the sign of the measured voltage changes accordingly which gives rise to an s-shaped or z-shaped anomalous Hall versus field trace. The sign of this curve (i.e. s-shaped or z-shaped) depends on the details of the material and the defect, as opposed to the ordinary Hall effect, which depends only on the sign (electron or hole) of the carriers (if there are multiple carrier types this is slightly more complicated). Skew scattering is predicted to have a well-defined scaling for the conductivities $\sigma_{xy,A} \propto \sigma_{xx}^n$, where $n=1$ for this specific case. In this regime side-jump scattering may also be active where the carriers scatter off defects, but the momentum direction is maintained via the carriers making a spatial jump transverse to the current direction. This results in a net Hall voltage transverse to the current. The side jump mechanism results in a scaling with $n = 0$.

For the lower conductivity regime (1 $M\Omega^{-1}\cdot cm^{-1} > \sigma_{xx} > 0.01$ $M\Omega^{-1}\cdot cm^{-1}$ or $1\mu\Omega\cdot cm < \rho_{xx} < 100 \mu\Omega\cdot cm$) intrinsic mechanisms dominate the buildup of charge transverse to the current direction. Intrinsic origins of the anomalous Hall effect date back to early proposals by Karplus-Luttinger(*16*). They proposed that an anomalous velocity develops in response to the internal magnetization. The group velocity of the electron wave packets contains a component that orients normal to the current, which cause a non-zero Hall voltage. This can be recast in terms of the Berry phase of the electronic band structure and therefore a purely quantum mechanical effect. Theory also predicts a well-defined scaling of $\sigma_{xy,A} \propto \sigma_{xx}^n$ with $n = 0$ in this lower conductivity regime. This coincides with the scaling of the side jump mechanism, which is detailed in Ref. (*15*). As with skew-scattering, the sign of the anomalous Hall voltage depends on a variety of factors of the electronic structure. So, the anomalous Hall versus field curves could be either s-shaped or z-shaped.

This basic information enables us to give a clear idea of which mechanisms active with increasing helium dosage in $PdCoO_2$ films. For low doses $PdCoO_2$ clearly sits in the high conductivity regime and exhibits a s-shape where the Hall resistance is positive for positively applied magnetic fields. The lack of hysteresis in this region shows that the helium dosing creates local moments, but those moments do not order. Increasing the dosage causes the emergence of another competing anomalous Hall mechanism which has the opposite sign, z-shape where the Hall resistance is negative for positive applied magnetic fields. The fact that there are two different shapes to the curves shows that there are two mechanisms. Finally, the resistivity increases (conductivity decreases) throughout the dosing process, nominally linearly with the dosage (see Fig. S14). This allows a systematic check of the theoretical scaling. Here, we find that there is a very interesting agreement with theory where the dependence of the Hall conductivity follows a linear dependence on conductivity with a non-zero intercept across all the conductivity regimes. Typical analysis of the anomalous Hall data utilizes a log-log plot where a rough scaling can be found within the specified regimes and the key message is that certain regimes are dominated by theorized mechanisms. The data here show that both mechanisms are active across the full conductivity range, which implies that these effects are additive across the regime probed. Interestingly, the emergence of the z-shaped response at higher dosages coincides with the onset of ferromagnetism which aids in identifying that this is an intrinsic effect.

### B. Interpretation of Muon Spectroscopy and Anomalous Hall Effect

Muon spectroscopy senses the long-range magnetic environment felt by an implanted muon: Experimentally, muons are implanted into a sample all with the same spin direction (for LE-μSR the spin of the muon is orthogonal to the momentum). If a sample is non-magnetic then the muons, now static in the sample, will precess in the plane of the detector array (see Fig. S7a) solely due to a small external magnetic field applied to the sample. This precession will occur all at the same rate prior to decay and measurement, which gives rise to large spin asymmetry (see Fig. S7 (b-c)). In a magnetic material, the local magnetic environment changes the spin precession, which appears in the spin asymmetry. For example, in polycrystalline Ni the spin asymmetry ($A_\mu$) is very small and inhomogeneity large ($\sigma$) due to strong local fields. For a ferromagnet to paramagnetic transition the spin-asymmetry will drop with reducing temperature, which indicates that the local magnetic field becomes non-zero. This is precisely what is seen here across 150 K with muons, which indicates that local ferromagnetic regions are forming. Moreover, at 25 K the kink and upturn likely indicates a substantial change in the long range magnetic ordering.

The anomalous Hall effect, however, shows that helium implantation causes ferromagnetic order that onsets near 25 K, which we assign as the temperature where hysteresis onsets. This is the point that long range magnetism is stabilized, and magnetization is locked in. This then manifests in transport which is sensitive to the local direction of the magnetization (i.e. the direction the current is deflected is dependent upon the local direction of the magnetization.). Above the Curie temperature there can still be an anomalous Hall effect, albeit without hysteresis (large kink in $R_{xy}$). This is attributed to the alignment of local moments in an applied magnetic field, which causes a large deflection of carriers transverse to the applied electric field. This can extend quite high above $T_C$, see below and Fig. S13 in the updated supplement. This could be due to local ferromagnetic regions embedded in a paramagnet (i.e. below the percolation threshold) or due to uncorrelated paramagnetic spins. Considering the muon results, the former is more likely. Overall, this is quite consistent with the physical scenario causing the emergence of magnetism. Specifically, helium implantation is random, and therefore will likely have lateral regions with higher local concentration and areas with lower concentrations. As such, it is perfectly reasonable to then guess that there is likely ordering up to 150 K, but below the percolation threshold. This then causes the strong anomalous Hall effect above $T_C$ (the temperature where hysteresis occurs) and also the non-zero volume fraction.

So, why, does the magnetic volume fraction never reach 100%? This likely comes from several reasons, (1) the sample thickness and (2) what each probe senses. First, the thickness of the samples used for muons is 50 nm compared to 10 nm used in the transport experiments. The larger thickness for muons was required to systematically probe the depth dependence of the magnetism (by changing the muon implantation energy, see Fig. 2) and the net signal (counts/second) in a muon experiment is quite low and a thicker sample simply yields a higher count rate. The magnetic volume fraction deviating from zero at 150 K definitively shows that local pockets of ferromagnetism are starting to form at that temperature. With reducing temperature more of these pockets form and enlarge until most of the sample becomes ferromagnetic at low temperature. As this sample is inhomogeneously dosed across the large thickness (implantation naturally has a gradient, see Fig. 2) the muons distribution always overlaps areas that are optimally dosed and areas that are under dosed, thus giving rise to something that is not 100% and should not be expected to be 100%.

## IV. Experimental Supporting Data

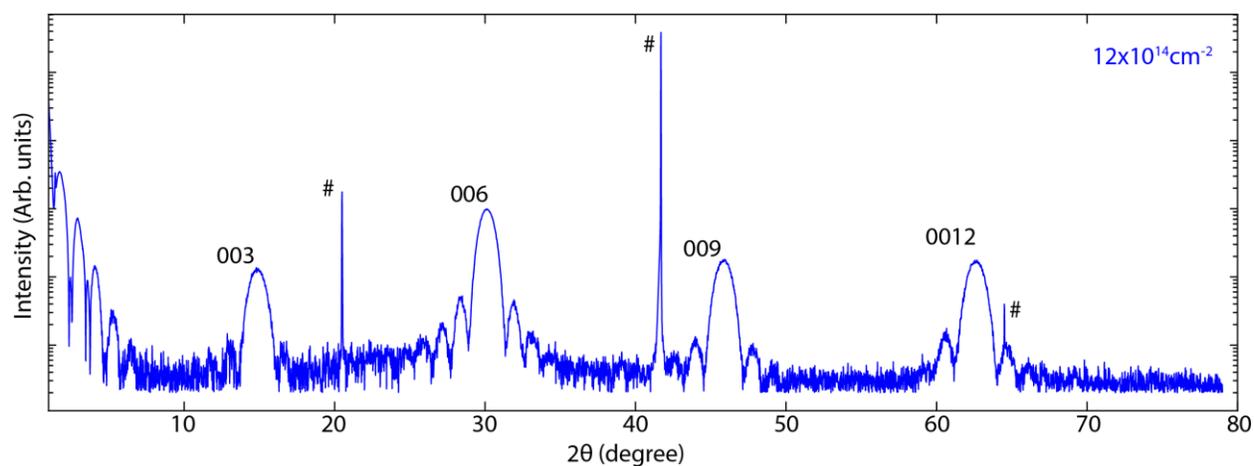

**Fig. S8**. Wide range $2\theta$-$\theta$ scans for a sample implanted to $12 \times 10^{14}$ cm$^{-2}$. The $00L$ PdCoO$_2$ peaks are labeled and the Al$_2$O$_3$ $00L$ peaks are indicated as # symbols. A background data set was subtracted to eliminate effects of low-angle air scattering.

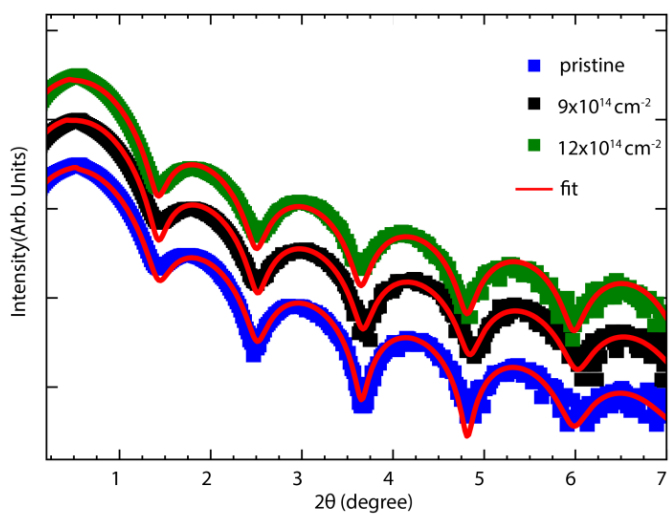

**Fig. S9**. $2\theta$-$\theta$ reflectivity scans for various doses. This shows the films surfaces are not damaged or changed by implantation.

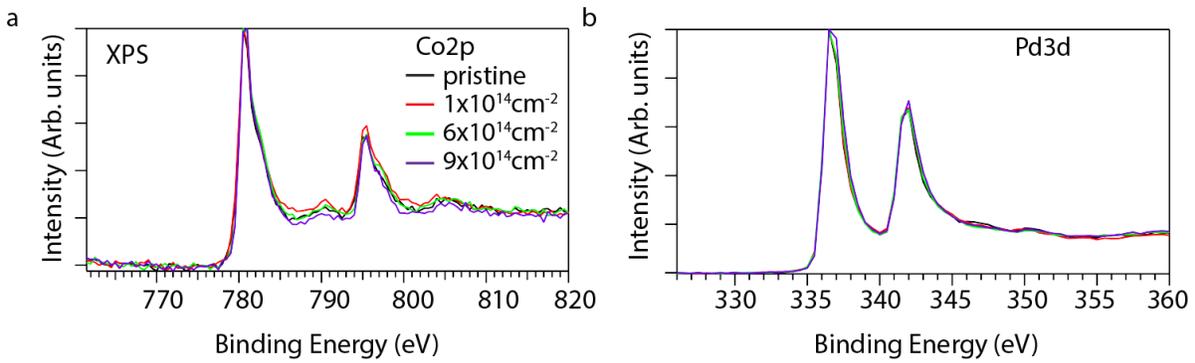

**Fig. S10** X-ray photoemission spectroscopy measured for the Co2p (a) and Pd3d b) edges for pristine and various dosages. The Co 2p states show a spin-orbit split doublet, with peaks at 780 eV and 796 eV, consistent with $Co^{3+}$ (*9, 17, 18*). Similarly, the Pd 2d shows two peaks at around 337 eV and 343 eV, consistent with Pd in a nominal +1 valence state, which matches well with $PdCoO_2$ (*9, 18*).

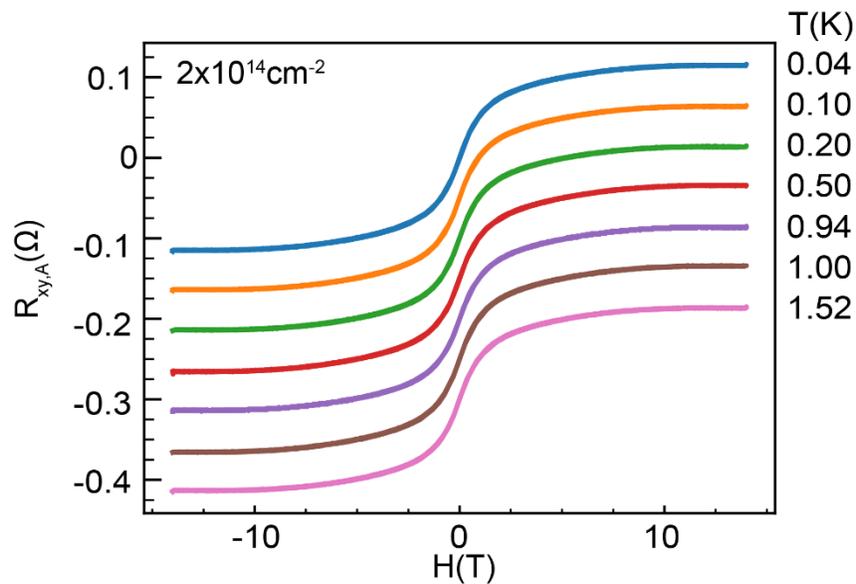

**Fig. S11**. Anomalous Hall resistance measured from 1.52 K down to 40 mK for a low dosage sample ($2\times10^{14}$ cm$^{-2}$) which shows no hysteresis.

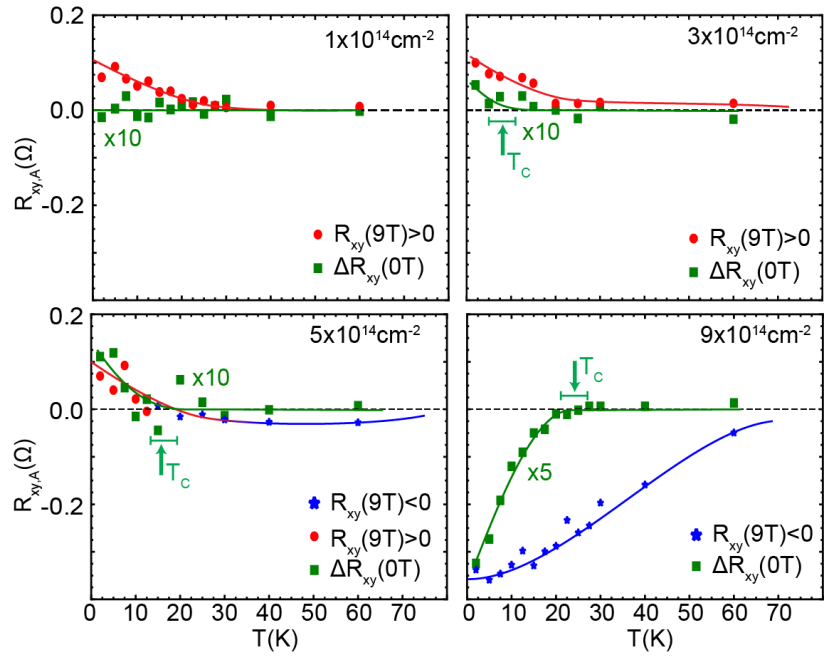

**Fig. S12**. Temperature dependence of the anomalous Hall resistance taken at $H=9$T, and $\Delta R_{xy,A}$ taken at zero magnetic field. The latter data is multiplied by the factor indicated for enhancement.

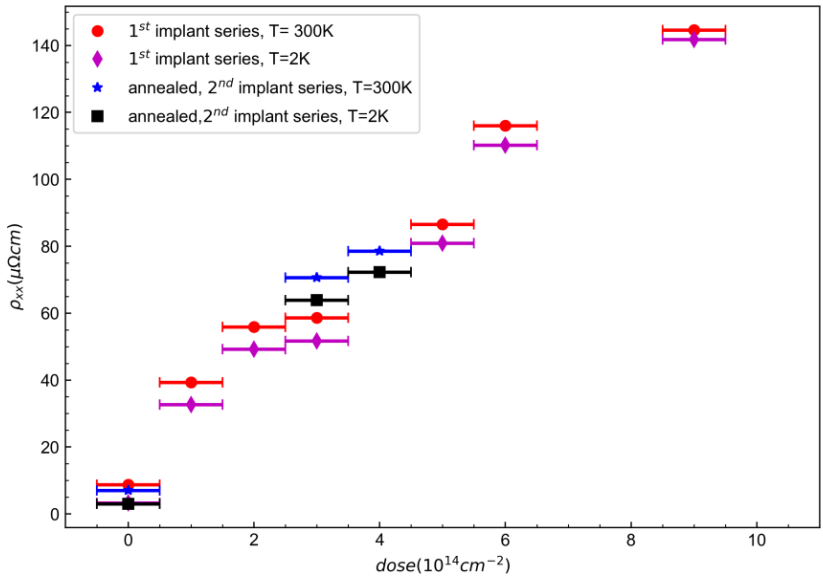

**Fig. S13**. Resistivity, $\rho_{xx}$, versus dose. The data shown here is for two implantation runs with increasing dose. The first run is shown as circles ($T = 300$ K) and diamonds (2 K) and the second run, following a reset anneal to 750 °C is shown as asterisks (300 K) and squares (2 K). This data shows that, within error of the dose (estimated $\sim\pm0.5\times10^{14}$ cm$^{-2}$), there an approximate linear relation among dose and resistivity. Moreover, this data shows that even at the highest dose ranges the resistivity is still about a factor of 10 lower than typical magnetic oxides.

**References**

1. F. Arnold, M. Naumann, H. Rosner, N. Kikugawa, D. Graf, L. Balicas, T. Terashima, S. Uji, H. Takatsu, S. Khim, A. P. Mackenzie, E. Hassinger, Fermi surface of PtCoO2 from quantum oscillations and electronic structure calculations. *Phys Rev B*. **101**, 195101 (2020).
2. J. P. Perdew, K. Burke, M. Ernzerhof, Generalized Gradient Approximation Made Simple. *Phys Rev Lett*. **77**, 3865–3868 (1996).
3. M. Casula, Beyond the locality approximation in the standard diffusion Monte Carlo method. *Phys Rev B*. **74**, 161102 (2006).
4. D. Staros, G. Hu, J. Tiihonen, R. Nanguneri, J. Krogel, M. C. Bennett, O. Heinonen, P. Ganesh, B. Rubenstein, A combined first principles study of the structural, magnetic, and phonon properties of monolayer CrI3. *J Chem Phys*. **156**, 014707 (2022).
5. C. A. Melton, L. Mitas, Many-body electronic structure of LaScO3 by real-space quantum Monte Carlo. *Phys Rev B*. **102**, 045103 (2020).
6. A. Annaberdiyev, C. A. Melton, G. Wang, L. Mitas, Electronic structure of αRuCl3 by fixed-node and fixed-phase diffusion Monte Carlo methods. *Phys Rev B*. **106**, 075127 (2022).
7. J. H. Skone, M. Govoni, G. Galli, Self-consistent hybrid functional for condensed systems. *Phys Rev B*. **89**, 195112 (2014).
8. D. G. Lock, V. H. C. Crisp, R. N. West, Positron annihilation and Fermi surface studies: a new approach. *Journal of Physics F: Metal Physics*. **3**, 561–570 (1973).
9. M. Brahlek, G. Rimal, J. M. Ok, D. Mukherjee, A. R. Mazza, Q. Lu, H. N. Lee, T. Z. Ward, R. R. Unocic, G. Eres, S. Oh, Growth of metallic delafossite PdCoO2 by molecular beam epitaxy. *Phys Rev Mater*. **3**, 093401 (2019).
10. E. Morenzoni, H. Glückler, T. Prokscha, H. P. Weber, E. M. Forgan, T. J. Jackson, H. Luetkens, C. Niedermayer, M. Pleines, M. Birke, A. Hofer, J. Litterst, T. Riseman, G. Schatz, Low-energy μSR at PSI: present and future. *Physica B Condens Matter*. **289–290**, 653–657 (2000).
11. T. Prokscha, E. Morenzoni, K. Deiters, F. Foroughi, D. George, R. Kobler, A. Suter, V. Vrankovic, The new beam at PSI: A hybrid-type large acceptance channel for the generation of a high intensity surface-muon beam. *Nucl Instrum Methods Phys Res A*. **595**, 317–331 (2008).
12. A. Suter, B. M. Wojek, Musrfit: A Free Platform-Independent Framework for μSR Data Analysis. *Phys Procedia*. **30**, 69–73 (2012).
13. E. Morenzoni, H. Glückler, T. Prokscha, R. Khasanov, H. Luetkens, M. Birke, E. M. Forgan, Ch. Niedermayer, M. Pleines, Implantation studies of keV positive muons in thin metallic layers. *Nucl Instrum Methods Phys Res B*. **192**, 254–266 (2002).
14. W. Eckstein, *Computer Simulation of Ion-Solid Interactions* (Springer Berlin Heidelberg, Berlin, Heidelberg, 1991), vol. 10.
15. N. Nagaosa, J. Sinova, S. Onoda, A. H. MacDonald, N. P. Ong, Anomalous Hall effect. *Rev Mod Phys*. **82**, 1539–1592 (2010).
16. R. Karplus, J. M. Luttinger, Hall Effect in Ferromagnetics. *Physical Review*. **95**, 1154–1160 (1954).
17. T. Burnus, Z. Hu, M. W. Haverkort, J. C. Cezar, D. Flahaut, V. Hardy, A. Maignan, N. B. Brookes, A. Tanaka, H. H. Hsieh, H.-J. Lin, C. T. Chen, L. H. Tjeng, Valence, spin, and orbital state of Co ions in one-dimensional Ca3Co2O6: An x-ray absorption and magnetic circular dichroism study. *Phys Rev B*. **74**, 245111 (2006).
18. H.-J. Noh, J. Jeong, J. Jeong, E.-J. Cho, S. B. Kim, K. Kim, B. I. Min, H.-D. Kim, Anisotropic Electric Conductivity of Delafossite PdCoO2 Studied by Angle-Resolved Photoemission Spectroscopy. *Phys Rev Lett*. **102**, 256404 (2009).